# Coexistence of Direct-to-Device (D2D) Satellite Systems with 5G/6G Terrestrial Networks: Cross-Border Aggregate Interference Analysis in Saudi Arabia

**MUNEER M. ALZUBI[1], MUHAMMAD ALRABEIAH[2], KHALID ALDUBAIKHY[3], ABDULLAH ALQASIR[3], MAJED ALSIFAEE[2], MOHAMED-SLIM ALOUINI[1], FELLOW**

[1]CEMSE, King Abdullah University of Science and Technology (KAUST), Thuwal 23955, Saudi Arabia
[2]Communications, Space & Technology Commission (CST), Riyadh 75606, Saudi Arabia
[3]Department of Electrical Engineering, Qassim University, Buraydah 52571, Saudi Arabia

CORRESPONDING AUTHOR: Muneer M. AlZubi (e-mail: Muneer.zubi@kaust.edu.sa).

This work is sponsored by the Communications, Space & Technology Commission (CST) of Saudi Arabia via the Spectrum Advisory Group (SAG).

**ABSTRACT** Direct-to-device (D2D) satellite systems, particularly those based on 3GPP-compliant non-terrestrial networks (NTN), deliver connectivity to standard mobile phones and IoT devices using terrestrial International Mobile Telecommunications (IMT) spectrum. This raises the risk of harmful cross-border interference into incumbent terrestrial IMT networks, an issue considered under Agenda Item 1.13 of the ITU World Radiocommunication Conference 2027 (WRC-27). This paper presents an ITU-R-compliant assessment of aggregate downlink interference from two hypothetical D2D systems operating in Jordan and Iraq into terrestrial IMT user equipment (UE) in Saudi Arabia, operating in the 698-960 MHz band. Unlike prior studies that evaluate a few test points near a simplified ideal tri-border intersection, we distribute multiple victim UEs along realistic national borders within various strips at a distance of 0-40 km from the borders. We compare two satellite selection strategies: highest-elevation and random strategies. The CDF of the interference-to-noise ratio (I/N) together with the intra-system and inter-system aggregation factors are analyzed. Results show that the protection criterion is exceeded up to 88% of the time, requiring up to 27 dB of additional isolation, so coexistence is infeasible without interference mitigation. In general, random selection produces higher interference because low-elevation beams place victims closer to boresight. Two mitigation techniques are then evaluated. A border isolation zone of one cell diameter yields only a moderate improvement under both strategies, whereas a narrow-beam, deep-null phased-array pattern lowers the exceedance to about 40% irrespective of strategy, providing an important input for spectrum regulatory considerations.

## I. INTRODUCTION

Over the last five years, satellite communications have undergone a major transformation, thanks to the rise of commercial direct-to-device (D2D) systems that use low Earth orbit (LEO) satellites to deliver voice, messaging, and broadband data to standard smartphones [1, 2]. This emerging technology uses terrestrial International Mobile Telecommunications (IMT) spectrum to deliver wide-area connectivity where terrestrial mobile networks are missing or too costly. The International Telecommunication Union (ITU) officially refers to it as direct connectivity between mobile satellite service and IMT user equipment (DC-MSS-IMT) [3]. This capability is enabled by the following technologies: LEO mega-constellations, 3GPP's standardization of Non-Terrestrial Networks (NTN), high-gain phased-array antennas on satellites, and advanced on-board signal processing. These advanced technologies make it feasible for LEO satellites to produce multiple narrow beams, each transporting 5G New Radio (NR) traffic to users within IMT frequency bands [2]. This technology can close digital divides in regions not covered by terrestrial networks, including rural, maritime, and underserved regions.

Despite the compelling promise of the D2D system, it presents significant regulatory and technical challenges. Many sub-3GHz bands are globally identified for IMT 4G/5G mobile networks in many countries. National regulators must balance the access of D2D systems to spectrum bands identified for IMT and the risk of causing harmful aggregate interference to IMT networks. The risk becomes particularly critical in border regions, as satellite beams authorized in one country inevitably extend their coverage into adjacent countries [4, 5]. Recognizing the urgency and complexity of this challenge, the 2023 World Radiocommunication Conference (WRC-23) adopted Resolution 253 [6]. This resolution formally directs ITU-R Working Party 4C (WP 4C) to carry out comprehensive sharing and compatibility studies for D2D across different IMT frequency bands, with results to guide decisions at WRC-27 Agenda Item 1.13 [3].

The coexistence of satellite-terrestrial networks, such as D2D systems and incumbent terrestrial IMT systems operating in co-channel or adjacent frequency bands, has been the subject of extensive study [1, 5, 7]. Early work focused on the uplink interference from terrestrial IMT networks into satellite receivers. In [8], the co-channel and out-of-band interference from 5G/6G terrestrial networks into satellite services is characterized to derive design criteria to protect satellite services. In [9], the authors constructed stochastic interference models for the uplink between terrestrial stations operating in the upper-6 GHz band and satellite receivers, with statistical treatments applied to both array antenna gain and clutter loss. In [10], a stochastic geometry-based analytical framework is used to examine the impact of terrestrial interference on the uplink of a LEO satellite constellation in high IMT bands, offering an analytical benchmark for validating simulation. A time-varying aggregate interference model for LEO satellite-terrestrial cellular coexistence is developed analytically, characterizing the interference via stochastic modelling of satellite mobility [11]. This model can be used to analyze interference caused by terrestrial cellular networks on satellite communication systems. In [12], the authors assess the uplink interference from terrestrial IMT networks into D2D satellite links using Monte Carlo simulation. This study concludes that coexistence is feasible under appropriate regulatory conditions.

However, the current primary concern is the downlink interference from D2D satellites into the terrestrial IMT networks as incumbent services, i.e., terrestrial base station (BS) and user equipment (UE). A number of sharing and compatibility studies on aggregate interference between D2D systems and terrestrial IMT networks have been carried out by ITU-R Working Party 4C (WP 4C) as contributions to WRC-27 Agenda Item 13. These studies can help in assessing to which degree D2D systems can coexist with mobile networks in frequency bands identified for IMT. In [13], several African countries conducted a study to evaluate cross-border aggregate interference at 700 MHz from two D2D systems serving Botswana and Zimbabwe into a single victim IMT UE located at the tri-border area inside South Africa. In this study, a multi-system aggregation factor (β) and an intra-system aggregation factor (α) are calculated, from which the power flux density (PFD) regulatory limits are derived based on IMT UE protection criteria. In [14, 15], studies conducted by China evaluate aggregate downlink interference from D2D systems into a single terrestrial IMT UE in the 1920-1980/2110-2170MHz bands across three tri-border scenarios. Considering that this study used the highest-elevation satellite selection strategy, it concludes that only satellite sidelobes contribute to cross-border interference into a neighboring country that does not authorize the D2D service. This study accounts for D2D intra-system beam isolation distances and inter-system geofencing to prevent mutual interference, concluding that co-frequency D2D systems from different operators can not simultaneously cover the same area and must maintain geographic separation. The co-channel and out-of-band interference from D2D systems into a single terrestrial IMT BS and UE operating in the 1920-1980MHz (uplink) and 2110-2170MHz (downlink) bands is evaluated using a cross-border scenario [16, 17]. In these studies, the D2D systems are assumed to operate in Brazil and Argentina, while the victim IMT network is located in Paraguay. The studies examine a beam-pointing exclusion zone along the border that directs satellite main beams into D2D operating countries as a mitigation technique that limits interference into Paraguay to sidelobe emissions only. In [18], Iran evaluates four interference scenarios into a single terrestrial IMT UE and BS in a neighboring country in the 1920-2170 MHz and 2500-2690 MHz bands using a deterministic single-satellite model. The results show that satellite downlink scenarios require a reduction in the effective isotropic radiated power (EIRP) to satisfy the protection criterion, while D2D UE uplink scenarios produce no interference, confirming that the uplink environment remains unchanged when a standard IMT UE is used. In [19], a study conducted in India simulates aggregate downlink interference from D2D into a fixed single IMT UE at 2 GHz. The study concludes that constellation density is the main driver and harmful interference can arise from non-bordering countries' systems. However, this study assumes that all visible satellites in the constellation contribute to interference without defining a realistic service area of the D2D system. France implemented a statistical cross-border study involving a France-Germany-Switzerland tri-border scenario at 2600 MHz [20]. It simulates a D2D system, assuming the IMT UEs are located at several points in the tri-border area to derive α and β aggregation factors. In [21, 22], the authors use Monte Carlo simulation to study cross-border interference between D2D systems and IMT networks at 2150 MHz and 1900 MHz, in accordance with IMT protection criteria. Results indicate that protecting both IMT UEs and BSs requires specific isolation distances as a mitigation technique, with the shorter isolation distance achieved via directional antennas and stricter elevation constraints. Most of the published work on tri-border interference coexistence analysis did not examine mitigation techniques to solve the coexistence issue and did not incorporate realistic cross-border scenarios into their analysis. In addition, they use either a single test point or a few points for the victim around the tri-border area only (e.g., IMT UE or BS). This may not represent the true worst case or actual interference along the overall country's borders.

In this paper, we analyze the cross-border aggregate downlink interference for D2D systems into the terrestrial IMT UEs in the 698-960 MHz band, focusing on a tri-country scenario. This is an extension of a study submitted to ITU-R Working

Party 4C (WP4C) by the Saudi Arabia administration [23]. The D2D systems are assumed to provide service coverage in Jordan and Iraq, which causes co-channel interference to victim IMT UEs located inside the border area of Saudi Arabia. The regulatory concern is that the LEO beam footprints span tens of kilometres, and thus satellites serving border cells in Jordan/Iraq can illuminate IMT UEs across the Saudi Arabia border at significant power levels even though Saudi Arabia is assumed not to have D2D service coverage. Moreover, we examine the impact of the isolation distance inside the D2D operating countries on the interference level in terms of interference-to-noise (I/N) ratio and cumulative distribution function (CDF). The main contributions of this paper are as follows:

1. We present, to the best of our knowledge, the first rigorous ITU-R compliant characterization of cross-border aggregate D2D downlink interference in the 698-960 MHz band into terrestrial IMT UEs. This also includes the Intra-system and inter-system aggregation factor analysis in the 698-960 MHz band.
2. Instead of studying the aggregate interference on a few victims located only in the tri-border area, 100 test points for victim IMT UEs are randomly distributed along the borders of Saudi Arabia to accurately examine the aggregate interference. In addition, we distributed the victim UEs into multiple strips (i.e., 100 UEs per strip) with distance from the border to examine the effect of distance on the overall interference.
3. Actual country borders are considered between Saudi Arabia, Jordan, and Iraq instead of simplified shapes. This affects the actual locations of D2D cells in the neighboring countries and victim UEs in the border region of Saudi Arabia, which leads to accurate aggregate interference analysis along the borders
4. We compare the aggregate D2D downlink interference levels using two satellite selection strategies: the random satellite selection strategy and the highest-elevation selection strategy. This helps provide a technical basis for operational constraints and optimum mitigating techniques that could be of great value to regulators.
5. We show that satellite selection strategy critically affects the spatial origin of cross-border interference: highest-elevation selection confines harmful interference to border-adjacent cells, while random selection extends it to border-distal cells, as lower elevation angles tilt the satellite beam sufficiently to illuminate victim UEs across the border even from distant service areas.
6. We evaluated two interference mitigation techniques, including satellite beam isolation distance and phased-array antenna radiation pattern, demonstrating that while both reduce aggregate interference at the IMT UE victim, neither alone is sufficient to meet the protection criterion, highlighting the need for additional or combined regulatory measures.
7. We obtain the CDF of aggregate interference and the additional reduction in interference power needed to satisfy the IMT protection criteria. This can translate the simulation results into actionable regulatory inputs.
8. We document a complete ITU-R-compliant simulation methodology integrating Simplified General Perturbations 4 (SGP4) orbital mechanics, antenna models, propagation models, and the Monte Carlo engine, establishing a replicable framework for D2D compatibility studies in other geographic regions, frequency bands, and constellation configurations.

The remainder of this paper is organized as follows. Section II describes the system model, including the D2D constellation and antenna parameters, the terrestrial IMT victim receiver, and the cross-border geographic scenario. The Monte Carlo simulation methodology and interference mitigation techniques are described in Section III. Section IV presents the simulation results for the individual D2D system and the combined two-system scenario with and without mitigation techniques. Section V concludes the paper.

## II. SYSTEM MODEL

In this section, we present the system model that describes the cross-border geographic scenario including the D2D satellite constellation and the terrestrial IMT UEs. The cross-border region of Saudi Arabia (country A), Jordan (country B), and Iraq (country C) presents a challenging environment in which the terrestrial IMT networks in Saudi Arabia may be exposed to interference from D2D systems providing service coverage in neighboring countries, as shown in Figs. 1 and 2. There is no D2D system providing service coverage in Saudi Arabia (country A). The terrestrial IMT UEs located in country A represent the victims that receive downlink aggregate interference from the D2D satellite constellations. The current study concerns the protection of IMT UEs in country A. The D2D systems and terrestrial IMT networks operate over the same frequency in the three countries. Each cell in countries B and C is covered by a single satellite and a single system. This introduces a new, transboundary interference vector that requires rigorous assessment. This study addresses a critical gap in spectrum coexistence analysis by quantifying the interference impact of D2D systems on IMT networks in a geographically sensitive tri-border region, considering realistic borders. In this study, we quantify the CDF of aggregate I/N ratios from D2D transmissions considering orbital characteristics, antenna patterns, and propagation conditions. The cumulative interference impact of two D2D systems providing service coverage areas in countries B and C is compared with the ITU protection threshold (-6dB [24]) to identify harmful interference scenarios. The interference level is calculated at several points distributed along the border region of country A at different distances from the borders, i.e., the five color strips shown in Fig. 2. In addition, this study computes the CDF of aggregation factors to quantify how interference from multiple satellites combines at the victim UE receiver. This includes the intra-system and inter-system aggregation factors. We also investigate the impact of the isolation distance for the D2D coverage area near the border area within countries B and C. This study supports regulatory coordination by contributing to spectrum regulatory considerations.

A thorough understanding of the technical and operational characteristics of the D2D systems and terrestrial IMT networks is essential for accurate interference analysis. The IMT UE characteristics used in this analysis are obtained from the Working Party 5D (WP 5D), as shown in Table 1 [25]. However, to account for worst-case interference, we assume the IMT UE antenna gain is equal to zero rather than -3dBi and body loss is equal to zero rather than 4dB values.

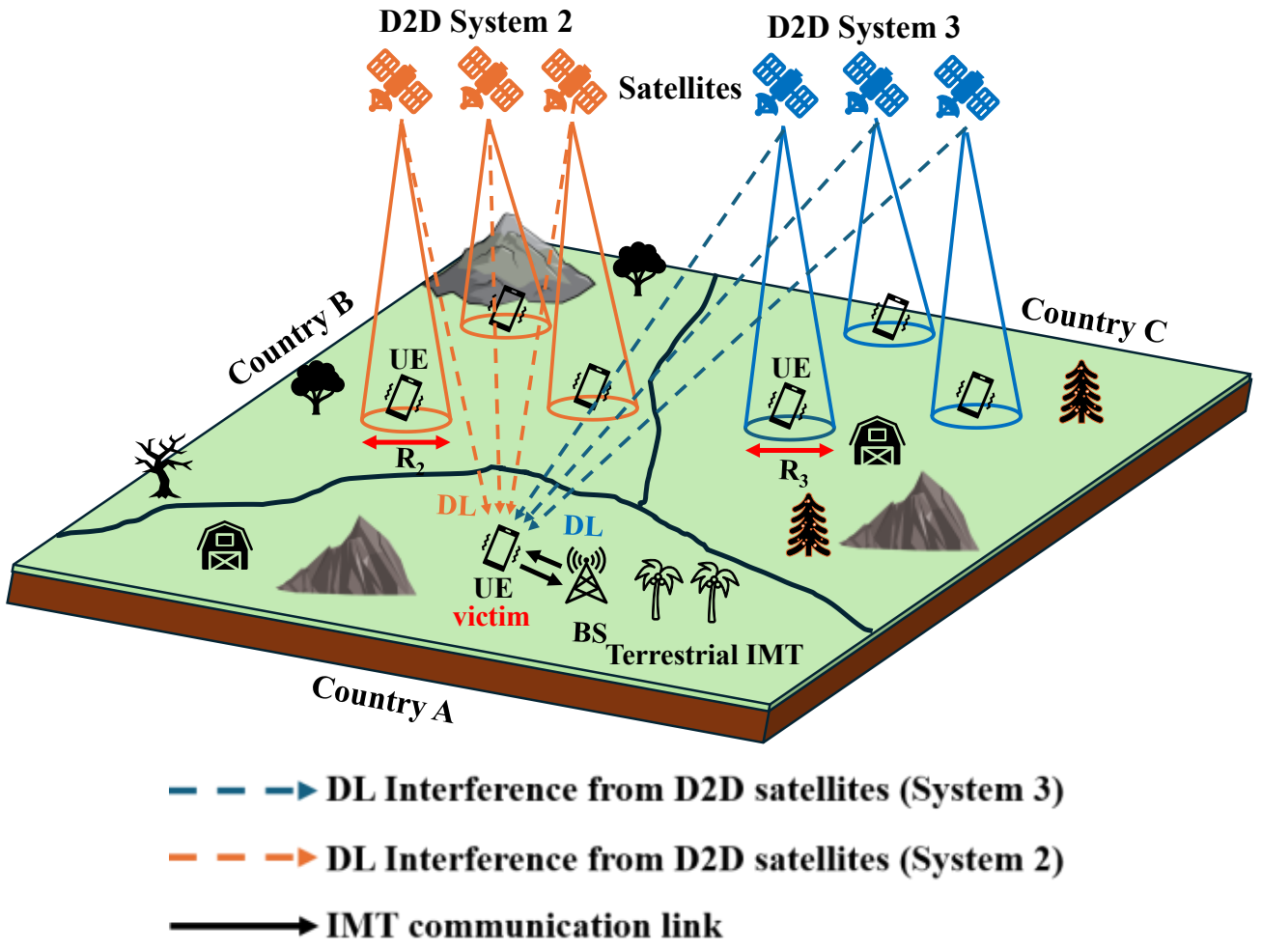

**FIGURE 1. Illustrative diagram of the cross-border interference scenario to protect terrestrial IMT UE in country A from two distinct D2D systems operating in neighbouring countries B and C.**

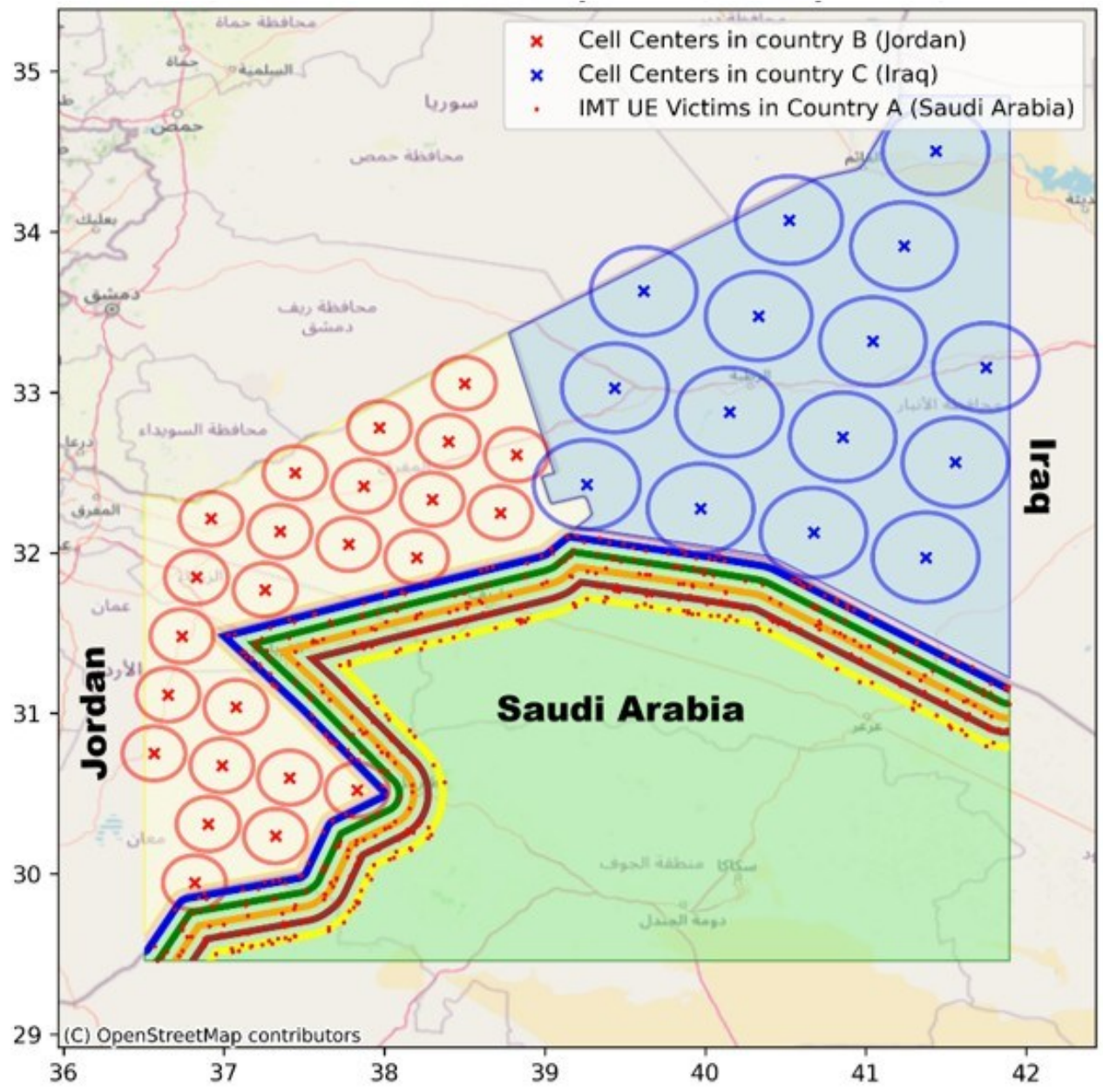

**FIGURE 2. Realistic simulation area for the cross-border interference scenario involving Saudi Arabia (country A), Jordan (country B) and Iraq (country C).**

The propagation pathloss between D2D satellites and IMT UEs on the ground is characterized following Recommendation ITU-R P.619-6 [26]. The IMT UEs are assumed to be near the border region inside Saudi Arabia, which is considered a rural area. Thus, the clutter loss can be ignored in these areas. Also, since the frequency under analysis is below 1 GHz, atmospheric attenuation can be ignored. Therefore, the pathloss model for the satellite-to-ground propagation channel can be reduced to the free space loss model.

**TABLE 1. Characteristics of terrestrial IMT UE.**

| Parameter | Unit | Value |
|---|---|---|
| Antenna pattern | - | Omni |
| Gain | dBi | 0 |
| NF | dB | 9 |
| Height | m | 1.5 |
| Body loss | dB | 0 |

**TABLE 2. Orbital parameters of D2D systems 2 and 3.**

| System ID | System 2 | System 3 |
|---|---|---|
| Frequency range (MHz) | 694/698-960 | 698-960 |
| Altitude (km) | 500 | 525 |
| Inclination (deg) | 55 | 53 |
| Number of planes | 60 | 28 |
| Satellites per plane | 60 | 120 |
| RAAN spacing (deg) | 6 | 12.9 |
| Total number of satellites | 3600 | 3360 |

In this study, two D2D LEO satellite constellation systems are considered, with orbital parameters listed in Table 2 [27]. Systems 2 and 3 refer to standardized reference LEO satellite constellation systems defined by ITU-R [27]. In this study, System 2 and System 3 are assumed to operate in countries B and C, respectively. The technical characteristics of D2D systems 2 and 3, listed in Tables 3 and 4, are used in the link budget calculation to evaluate the aggregate interference level from the D2D satellites into the terrestrial IMT UEs. The antenna radiation pattern of systems 2 and 3 follows the Recommendations ITU-R M.2101 and ITU-R S.1528 (1.4), respectively [27-29]. The normalized radiation patterns are plotted in Fig. 3 using the parameters listed in Table 4, which are provided in the technical and operational characteristics [27, Annex 1]. According to the ITU-R recommendation, the radiation pattern of System 3 using the parameters listed in Table 4 is considered applicable across all frequencies, i.e., fixed wavelength. The antenna radiation pattern is used in the interference link budget calculation and determines the D2D cell radius in countries B and C, i.e., the satellite footprint on the ground. We use the off-axis angle at the -7 dB level to obtain the ground footprint radius (R) using Eq. (1). The angles are equal to 2.5° (R=22km) and 4° (R=37km) for systems 2 and 3, respectively.

$$R = h\ \tan\theta_0 \qquad (1)$$

where $\theta_0$ is the -7dB half-beamwidth angle in (rad), and $h$ is the satellite altitude in (km).

The satellite constellation simulation employs the Skyfield astronomy library and the Simplified General Perturbations 4 (SGP4) propagator in Python. Skyfield provides a high-level tool for generating precise positions of LEO satellites. Its functionality is built on the SGP4 model, a widely used standard mathematical model for accurately predicting satellite position and velocity over time, based on Two-Line Element sets (TLEs). TLEs are the standard format for representing satellite orbital parameters. The D2D satellite constellations of Systems 2 and 3 are modelled by generating TLE data to represent the Walker Delta constellation pattern based on the orbital parameters defined in Table 2. Then, the positions of the satellites are estimated at each time instance for use in the Monte Carlo simulation of aggregate interference, as will be demonstrated in the next section. The orbital periods (T) of systems 2 and 3 are 94.6 minutes and 95.1 minutes, respectively.

**TABLE 3. Technical characteristics of D2D systems 2 and 3.**

| System ID | System 2 | System 3 |
|---|---|---|
| Minimum elevation (deg) | 35 | 20 |
| Antenna polarization | - | RHCP |
| Emission Bandwidth (MHz) | 5 | 5 |
| Max EIRP Spectral Density (dBW/Hz) | -27.1 | -28.1 |
| Maximum PFD on the ground per satellite (dBW/m$^2$/MHz) | - | -93.4 |
| Antenna pattern | Rec. ITU-R M.2101 [28] | Rec. 1.4 of ITU-R S.1528 [29] |

**TABLE 4. Antenna parameters of D2D systems 2 and 3.**

| System ID | Parameters | Unit | Value |
|---|---|---|---|
| System 2 | Single-element antenna gain | dBi | 2 |
| | Antenna array configuration (Row×Column) | - | 28×28 |
| | Horizontal 3dB beamwidth of single element | deg | 120 |
| | Vertical 3dB beamwidth of single element | deg | 120 |
| | Horizontal radiating element spacing | dH/λ | 0.5 |
| | Vertical radiating element spacing | dV/λ | 0.5 |
| | Horizontal Front-to-back ratio | dB | 30 |
| | Vertical Front-to-back ratio | dB | 30 |
| System 3 | Peak antenna gain (G_max ) | dBi | 34.1 |
| | radial and transverse sizes of the effective radiating area of the satellite transmits antenna (L_r = L_t) | m | 1.6 |
| | side-lobe ratio (SLR) | dB | 20 |
| | number of secondary lobes | - | 2 |
| | Wavelength (λ) | m | 0.15 |

## III. METHODOLOGY

This section presents the Monte Carlo simulation methodology employed to evaluate the aggregate interference from D2D satellite systems into victim IMT UEs near the cross-border area of Saudi Arabia, as shown in Fig. 2. The methodology is developed based on the frameworks established in ITU-R Working Party 4C contributions, particularly those documented in the working document for WRC-27 Agenda Item 1.13. In addition, we evaluated the efficacy of two mitigation techniques to reduce the interference into the terrestrial IMT UEs, including an isolation zone and phased array antenna radiation pattern.

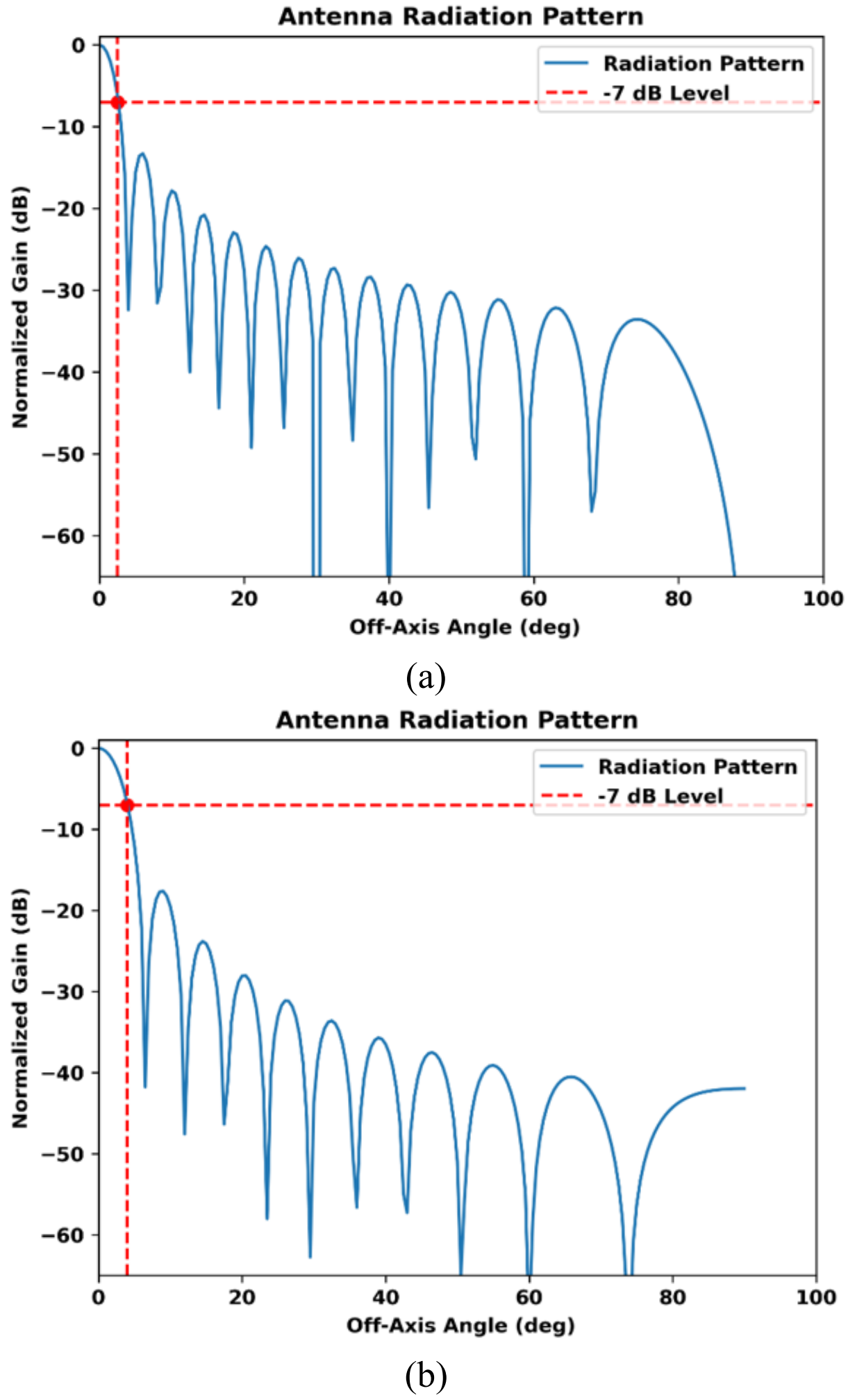

**FIGURE 3. Normalised radiation pattern as a function of boresight off-axis angle for the reference antenna of (a) the D2D System 2 given in Rec. ITU-R M.2101 and (b) the D2D System 3 given in Rec. 1.4 of ITU-R S.1528.**

### A. MONTE CARLO SIMULATION

The simulation is executed according to the following main steps:

**Step 1**: Define Simulation Area, D2D Ground Cells and IMT UE Locations.

The simulation area is defined to encompass the tri-border region of Saudi Arabia, Jordan, and Iraq, using realistic borders, as shown in Fig. 2. The simulation area dimensions are 600km×600km, with the approximate center of the tri-border region at the coordinate (32°09'13.9"N, 39°12'03.3"E). Service coverage of D2D systems in countries B and C is modelled using non-overlapping circular cells with the footprint radii of the corresponding system, calculated from the antenna radiation patterns as described in Section II. The victim IMT UEs are deployed within Saudi Arabia, with a focus on the border region to capture worst-case interference scenarios. A total of 100 victim IMT UEs per strip are distributed along the border within each strip of 5 km width, evaluated at varying distances from the border, i.e., d = 0-40 km.

**Step 2**: Generate D2D Satellite Constellation.

For each satellite in the constellation, the geographic position at each time step ($t$) is calculated using the Skyfield astronomy library and SGP4 propagator as described in Section II. The position of satellites is expressed in the Earth-Centered, Earth-Fixed (ECEF) frame, i.e., (X, Y, Z).

**Step 3**: D2D Satellite Selection and Beam Pointing Strategy.

A list of all visible satellites for each ground cell is obtained. The satellite is considered visible for any cell if the elevation angle ($\theta_e$) seen from the cell centre is larger than the minimum elevation angle ($\theta_{e,min}$) of the corresponding D2D system as given in Table 3, i.e., 35° for System 2 and 20° for System 3.

$$\theta_e = \sin^{-1}\left(\frac{(\boldsymbol{s}-\boldsymbol{c}).\boldsymbol{c}}{\|\boldsymbol{s}-\boldsymbol{c}\| \times \|\boldsymbol{c}\|}\right) \times \frac{180}{\pi} \quad (2)$$

where $\theta_e$ is the elevation angle in degrees, and $\boldsymbol{s}$ and $\boldsymbol{c}$ are positions of the satellite and cell center, respectively, in the ECEF frame.

Here, we assume each ground cell is served by a single satellite and a single system at a time. In this study, we examine the following satellite selection strategies to allocate satellites over these ground cells:

1. **Highest Elevation Satellite Selection**: For each cell, a satellite with the greatest elevation angle as seen from the cell's center is chosen from the list of visible satellites.
2. **Random Satellite Selection Strategy**: Each cell is assigned to a randomly selected satellite from the set of visible satellites, i.e., those above the minimum elevation angle.

The selected satellites point their main beams directly to the center of the assigned cells. The vectors between the satellites and the assigned cells are recorded ($\boldsymbol{v}_1$). Here, the position of ground cells is converted from the geodetic coordinate system (latitude, longitude, altitude) to the ECEF frame. A load factor of 100% is considered for the satellite systems in this simulation, i.e., all the assigned satellites are active and transmitting.

**Step 4**: Find the Satellite Off-Axis Angles ($\theta$) Toward the IMT UEs.

Find the angle between the satellite's main beam pointing direction (toward its serving cell) and the direction from the satellite to a victim IMT UE. This represents the angle between the two vectors: the first vector (satellite to cell center, $\boldsymbol{v}_1$) and the second vector (satellite to UE, $\boldsymbol{v}_2$). This angle is a critical parameter in interference calculations because it determines the satellite antenna gain in the direction of the victim receiver through the antenna radiation pattern.

$$\theta_{off} = \cos^{-1}\left(\frac{\boldsymbol{v}_1.\boldsymbol{v}_2}{\|\boldsymbol{v}_1\| \times \|\boldsymbol{v}_2\|}\right) \times \frac{180}{\pi} \quad (3)$$

where $\theta_{off}$ is the off-axis angle in degrees, $\boldsymbol{v}_1$ is the vector from the satellite to the cell center, and $\boldsymbol{v}_2$ is the vector from the satellite to the UE in the ECEF frame.

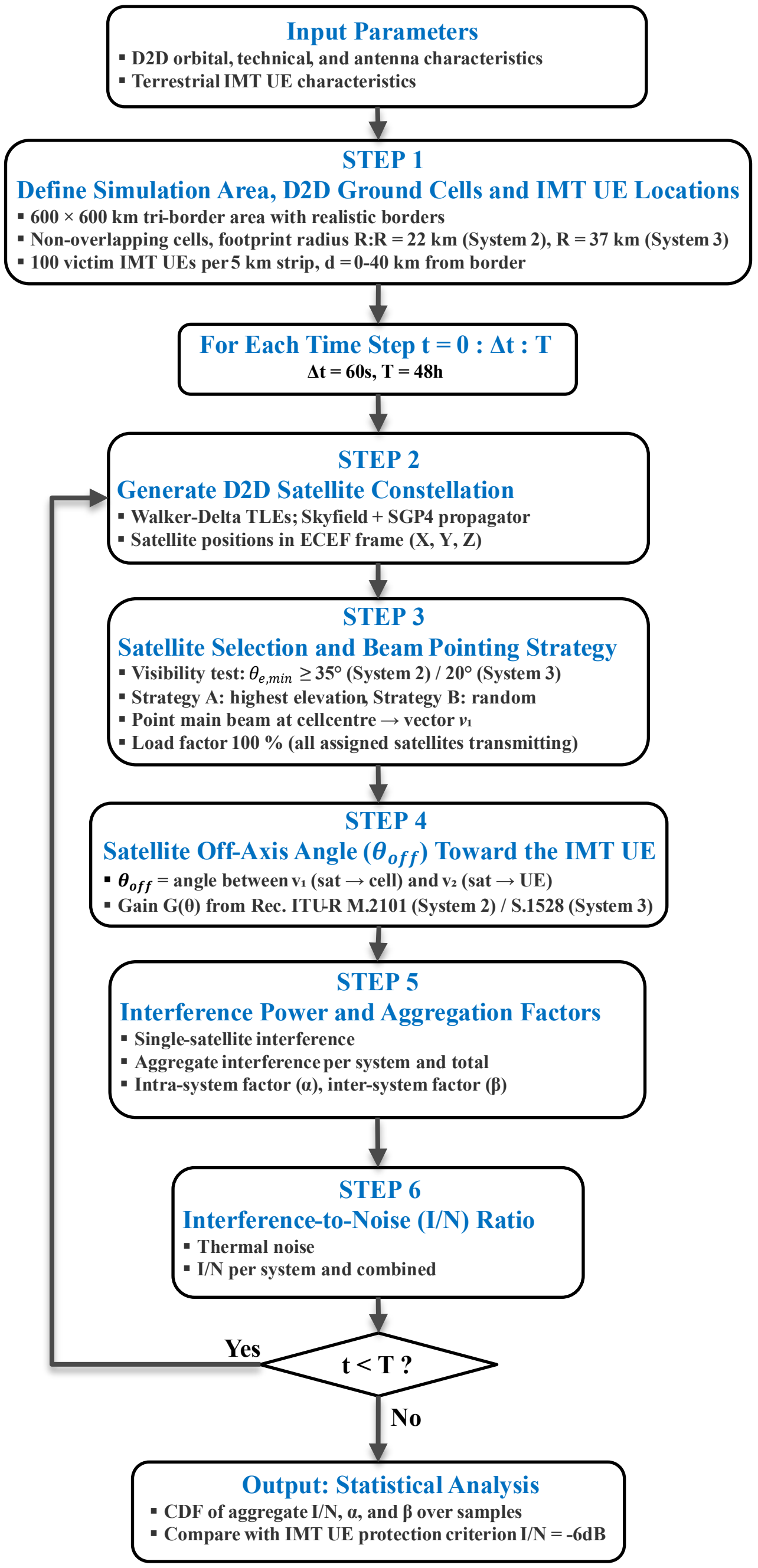

**FIGURE 4. Flowchart of the Monte Carlo simulation methodology used in this work.**

**Step 5**: Aggregate Interference and Aggregation Factors.

In this study, we assume that satellite EIRP is fixed and thus the PFD on the ground will change with the slant path and elevation Angle. The interference power from a single D2D satellite to a victim IMT UE is calculated at each time step as [14, 15]:

$$I_{m,n} = PSD_{dBw/Hz} + 10\,log_{10}(B) + G_{UE} + G_{sat}(\theta_{off}) - PL(r,f) - L_{body} \quad (4)$$

where the parameters in (4) are defined as follows:

$I_{m,n}$ : Interference power from the n-th satellite in the m-th D2D system in dBW.

$G_{sat}(\theta_{off})$ : Normalized gain of D2D satellite antenna in the direction of victim UE in dBi. It has a maximum value equal to 0 dBi and is calculated for each D2D system using the radiation pattern defined in Tables 3 and 4.

$PSD_{dBw/Hz}$ : Maximum EIRP spectral density of D2D system given in Table 3 in dBW/Hz.

$B$ : The channel bandwidth in Hz.

$PL(r, f)$ : Free-space path loss between the satellite and victim UE in dB.

$r$ : Slant distance between the satellite and victim UE in km.

$f$ : The operating frequency in MHz. In this study, $f$ = 800 MHz.

$G_{UE}$ : IMT UE antenna gain in dBi.

$L_{body}$ : Body loss for the victim UE in dB.

The aggregate interference level at each victim UE from all active visible satellites in each D2D system at each time step can be expressed in watts as [16, 17]

$$I_{system,m} = \sum_{n}^{N} 10^{I_{m,n}/10} \tag{5}$$

The total aggregate interference from both Systems 2 and 3 in watts at each time step can be calculated as

$$I_{total} = I_{system,2} + I_{system,3} \tag{6}$$

where $I_{system\ m}$ is the aggregate interference from the D2D system $m$ in watts; the index $m$ refers to the D2D system, i.e., m = {2, 3}, $I_{total}$ is the total aggregate interference from both Systems 2 and 3 in watts, and $N$ is the number of visible active satellites seen by the IMT UE.

The aggregation factor $\alpha$ is evaluated at the level of a single D2D system, from the interference powers received from different satellites of the same D2D system. It helps to understand how the interference power is distributed and whether there is a dominant interference direction. The factor $\alpha$ for the D2D system, $m$, is defined as [13, 20]:

$$\alpha = 10log_{10}\left(\frac{I_{\text{system,m}}}{I_{\text{max}}}\right) \tag{7}$$

where $I_{\text{max}} = \max(I_{\text{system,m}})$ is the highest interference received from a visible satellite in the D2D system $m$.

A second factor $\beta$ is defined to measure the apportionment/aggregation observed in country A from both D2D systems [13, 20]:

$$\beta = 10\log_{10}\left(\frac{I_{\text{total}}}{\max(I_{\text{system,2}}, I_{\text{system,3}})}\right) \tag{8}$$

**Step 6**: Interference-to-Noise Ratio (I/N) Calculation

The thermal noise power (dBW/Hz) for the victim IMT UE is calculated as [16]:

$$N = 10\ log_{10}(k \times T_0 \times B) + NF \tag{9}$$

where $k$ is Boltzmann's constant = 1.38 x$10^{-23}$ J/K, $T_0$ is the reference temperature = 290K, $B$ is the reference bandwidth in Hz, and $NF$ is the noise figure of the IMT UE in dB.

The thermal noise power in Eq. (9) can be expressed in dB as:

$$N = -204 + 10\ log_{10}(B) + NF \tag{10}$$

The I/N ratio at each victim UE from all active visible satellites in each D2D system at each time step can be expressed in dB as

$$\left(\frac{I}{N}\right)_m = I_{system,m}(\text{dB}) - N \tag{11}$$

$$I_{system,m}(\text{dB}) = 10\ log_{10}(I_{system,m}) \tag{12}$$

The I/N ratio at each victim UE from all active visible satellites in both D2D Systems 2 and 3 at each time step is given as

$$\left(\frac{I}{N}\right)_{total} = I_{total}(\text{dB}) - N \tag{13}$$

$$I_{total}(\text{dB}) = 10\ log_{10}(I_{total}) \tag{14}$$

The protection criteria for the IMT UE are expressed in terms of the I/N ratio at -6 dB. The simulation runs over a period of two days (48 hours) with a 60s time step for 100 UE locations per strip. Thus, we have a total of 288,000 samples. The simulation steps 2-6 are repeated at each time step.

### B. INTERFERENCE MITIGATION TECHNIQUES

Two mitigation techniques are evaluated in this study to reduce aggregate interference from D2D satellite downlinks into terrestrial IMT UEs along the border area.

1. **Isolation Distance:** the isolation distance defines a geographic exclusion zone along the international border within which the D2D system is prohibited from directing satellite beams. Any D2D UE served by the constellation must be located at a minimum horizontal distance $d_{\text{iso}}$ from the border, ensuring that no satellite beam is pointed toward the protected zone. This is enforced by beam-pointing at the network operations center (NOC), such that the serving satellite selects only beams whose footprint centers fall outside the exclusion area. Increasing isolation distance reduces the number of active satellite beams that can interfere with the victim IMT UE near the border, thereby reducing the aggregate interference level. The isolation distance can be evaluated as a varying parameter to identify the minimum value that could satisfy the IMT UE protection criterion of $\frac{I}{N} \leq -6$dB.

2. **Phased Array Antenna Radiation Pattern:** the radiation pattern of the satellite transmit antenna governs the spatial distribution of radiated power and determines the off-axis interference received by terrestrial UE victims. For D2D systems employing phased-array antennas, the radiation pattern is modelled in accordance with Rec. ITU-R M.2101, which defines a composite pattern accounting for array element gain, beamforming gain, and sidelobe roll-off. Compared to the aperture antenna model of Rec. ITU-R S.1528 used for reflector antenna-based systems, the phased-array pattern exhibits steeper sidelobe attenuation away from the main beam boresight as shown in Fig. 5. Since the victim IMT UE in the cross-border scenario is outside the intended service area of the D2D beam, it lies in the sidelobe region of the satellite antenna, particularly when the maximum elevation satellite selection strategy is applied. The phased-array pattern thus provides inherent geometric mitigation where the deeper the sidelobe suppression at the off-pointing angles, the lower the aggregate interference at the victim UE.

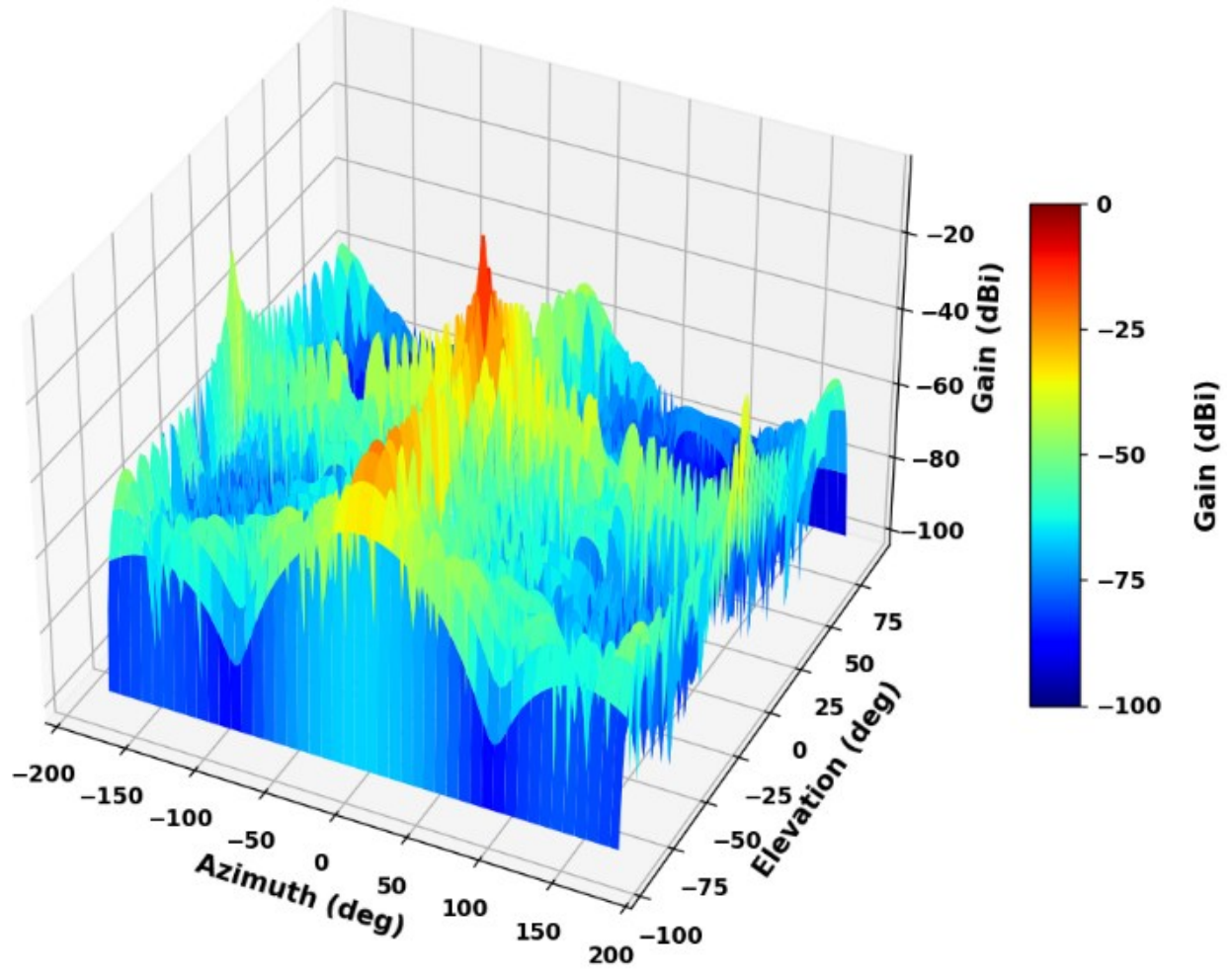

(a)

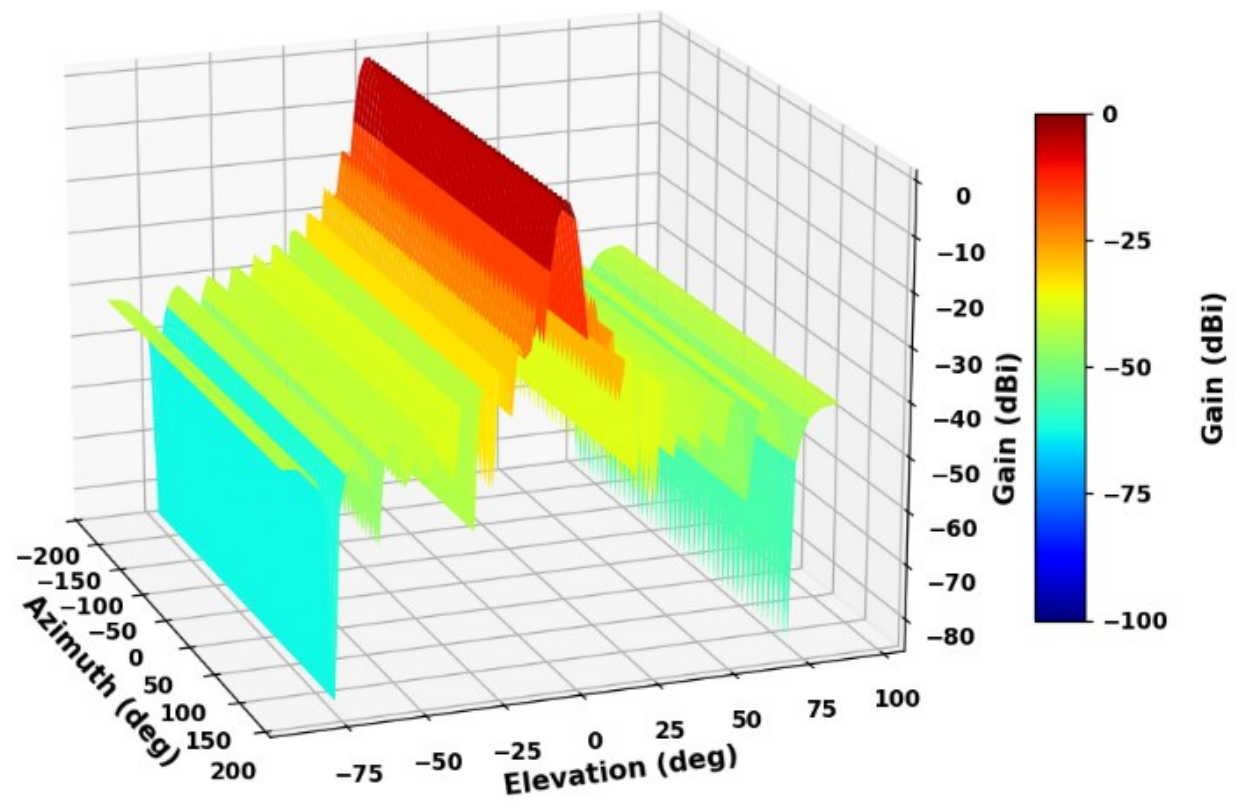

(b)

**FIGURE 5. Normalized 3D radiation pattern as a function of elevation and azimuth angles for the reference antenna of (a) Rec. ITU-R M.2101 and (b) Recs. 1.4 of ITU-R S.1528.**

## IV. RESULTS AND DISCUSSION

In this section, we discuss and analyze the obtained CDFs of the aggregate I/N ratio and the intra-system/inter-system aggregation factors for both highest-elevation and random satellite selection strategies. In addition, we evaluated two mitigation techniques to reduce the aggregate interference into the victim UEs.

### A. AGGREGATE I/N RATIO WITH HIGHEST ELEVATION AND RANDOM SATELLITE SELECTION STRATEGIES

The CDF of the aggregate I/N ratio of D2D Systems 2 and 3 for the victim UE points distributed in different strips at a distance (d=0-40km) from the border region of country A is shown in Figs. 6 and 7, respectively. These figures present the CDF of the aggregate I/N ratio resulting from interference generated by each D2D system.

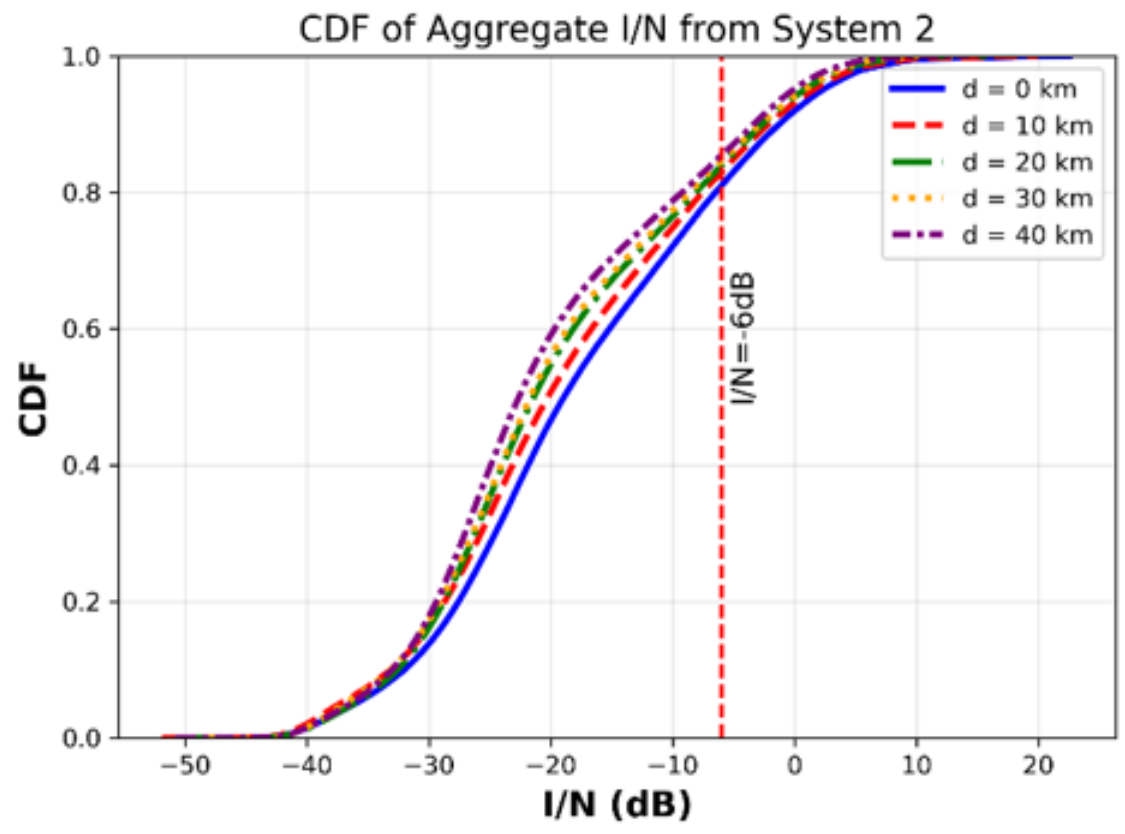

(a)

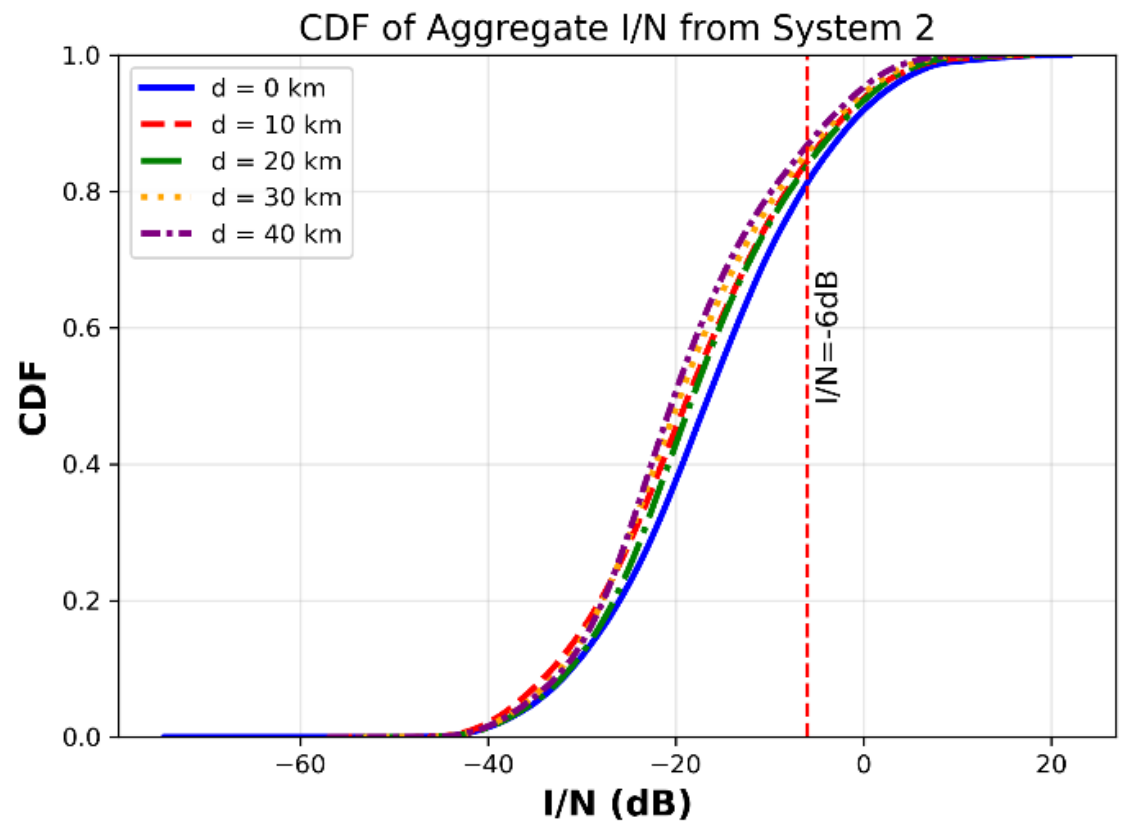

(b)

**FIGURE 6. The CDF of the aggregate I/N ratio of the D2D System 2 at different distances from the border region of country A for (a) highest elevation satellite selection strategy and (b) random satellite selection strategy.**

Results show that the IMT UE protection criterion (I/N=-6dB) is exceeded by the D2D System 2 for all distances, with about 18% exceedance of CDF, as shown in Fig. 6. On the other hand, the IMT UE protection criterion from the D2D System 3 is exceeded by about 81% of the CDF under the highest-elevation satellite selection strategy and by about 88% under the random satellite selection strategy, as shown in Fig. 7. Unlike the smooth CDF obtained for System 2, the CDF curves of System 3 exhibit a stepped shape. This comes from the antenna radiation pattern rather than the constellation geometry. Both patterns are fixed functions of the off-axis angle, so the time variation of the interference arises only from the changing geometry sweeping that angle across a static curve. System 3 uses the radiation pattern of Rec. ITU-R S.1528 (1.4) with a constant far-sidelobe floor as shown in Fig. 5(b), which yields a small number of discrete gain levels and hence clustered interference values. On the other hand, the phased array radiation of System 2 (Rec. ITU-R M.2101) has closely spaced nulls and sidelobes that sample an almost continuous range of gain as shown in Fig. 5(a). This is demonstrated later in this section, where replacing the System 3 antenna pattern with the phased array pattern removes the steps and provides a smooth CDF curve.

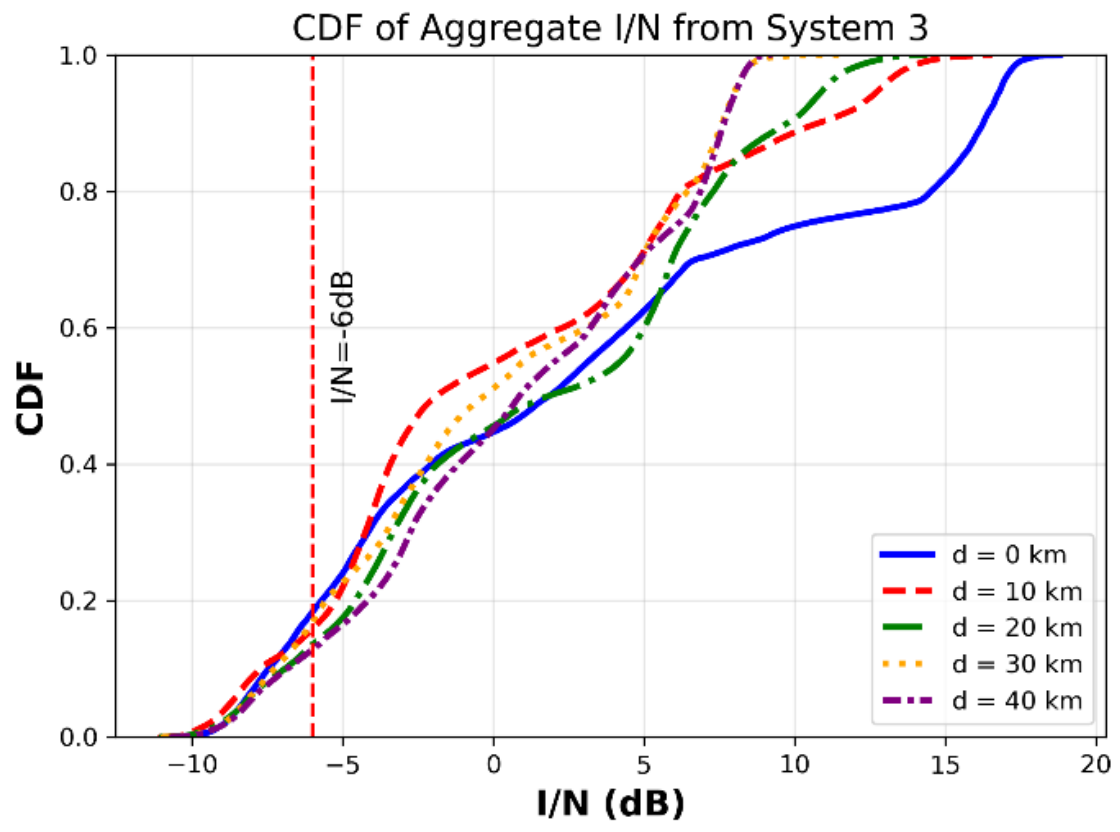

(a)

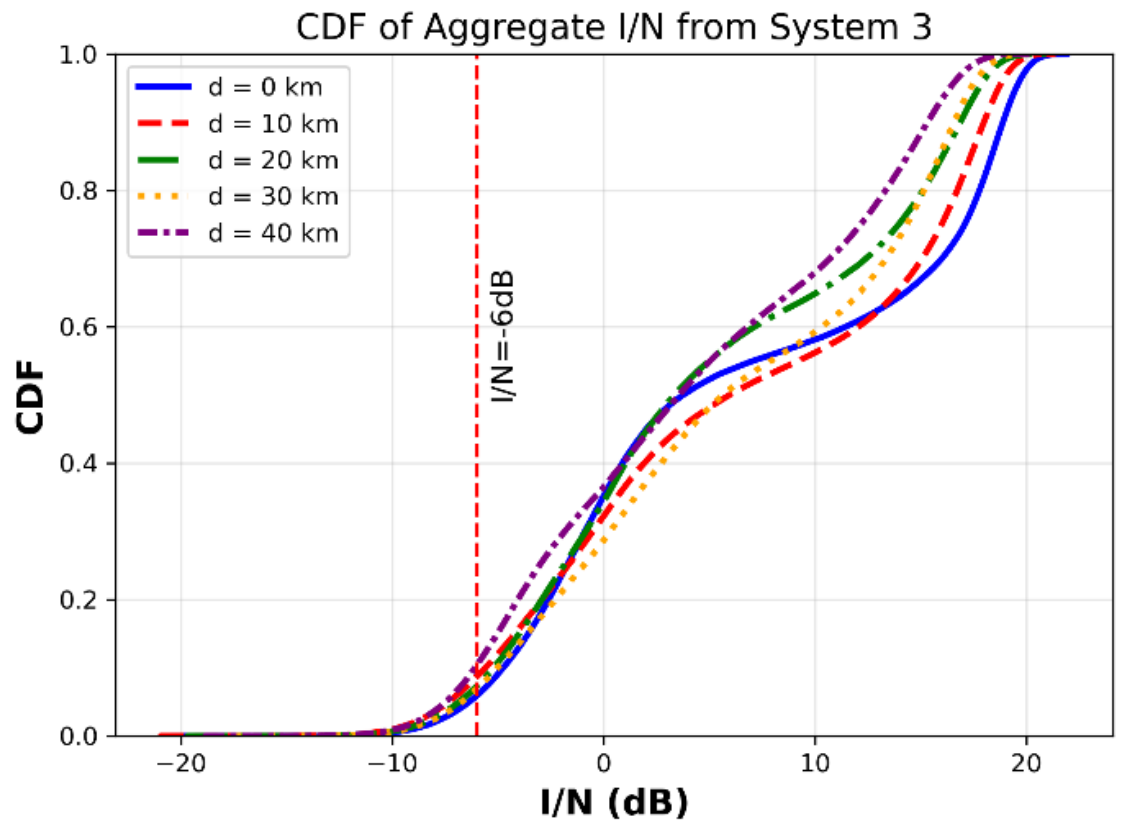

(b)

**FIGURE 7. The CDF of the aggregate I/N ratio of the D2D system 3 at different distances from the border region of country A for (a) highest elevation satellite selection strategy and (b) random satellite selection strategy.**

In both figures, the strip closest to the border (d=0km) produces the longest upper tail, confirming that victims immediately adjacent to the border experience the most severe worst-case interference. The interference effect of System 2 is less than that caused by System 3, and thus, the total interference is dominated by System 3, as shown in Fig. 8. Figure 8 presents the CDF of the aggregate I/N ratio resulting from interference generated by both D2D systems. The reason for the lower interference caused by System 2 is due to the phased-array antenna radiation pattern of System 2, which has a narrower beam with multiple deep nulls between sidelobes that reduce the gain and thus the interference toward the IMT UE victims.

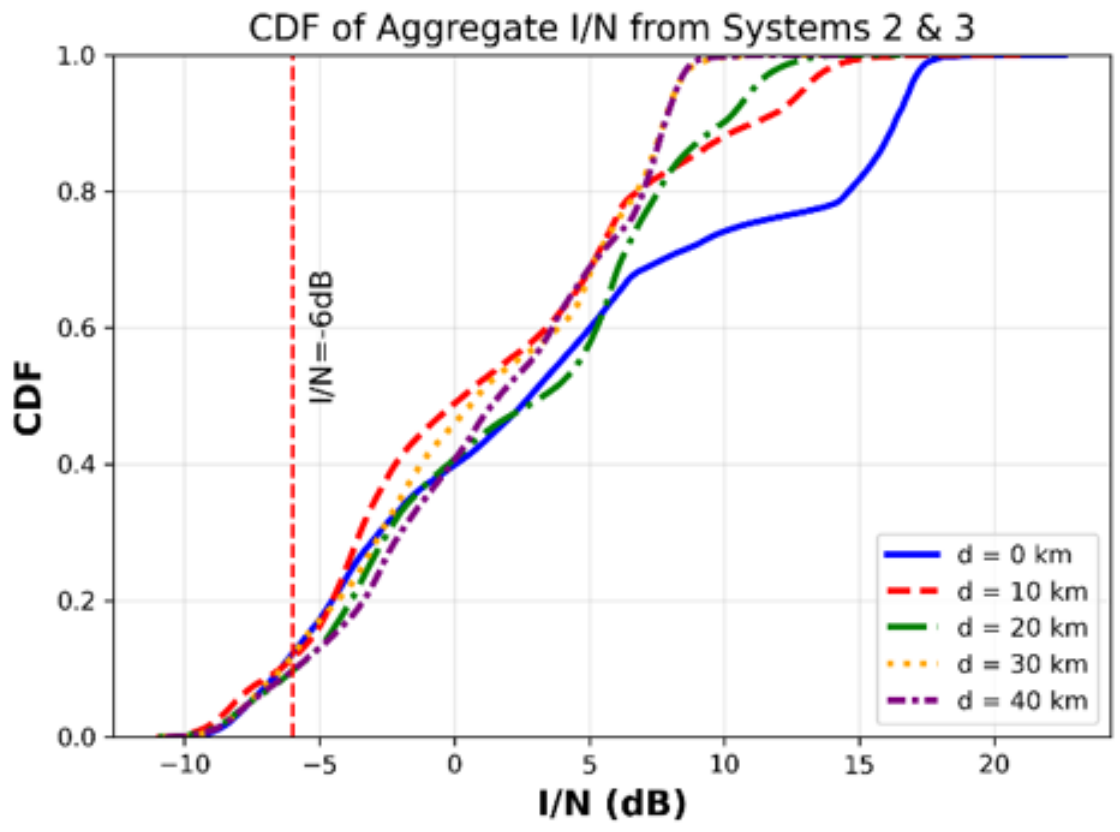

(a)

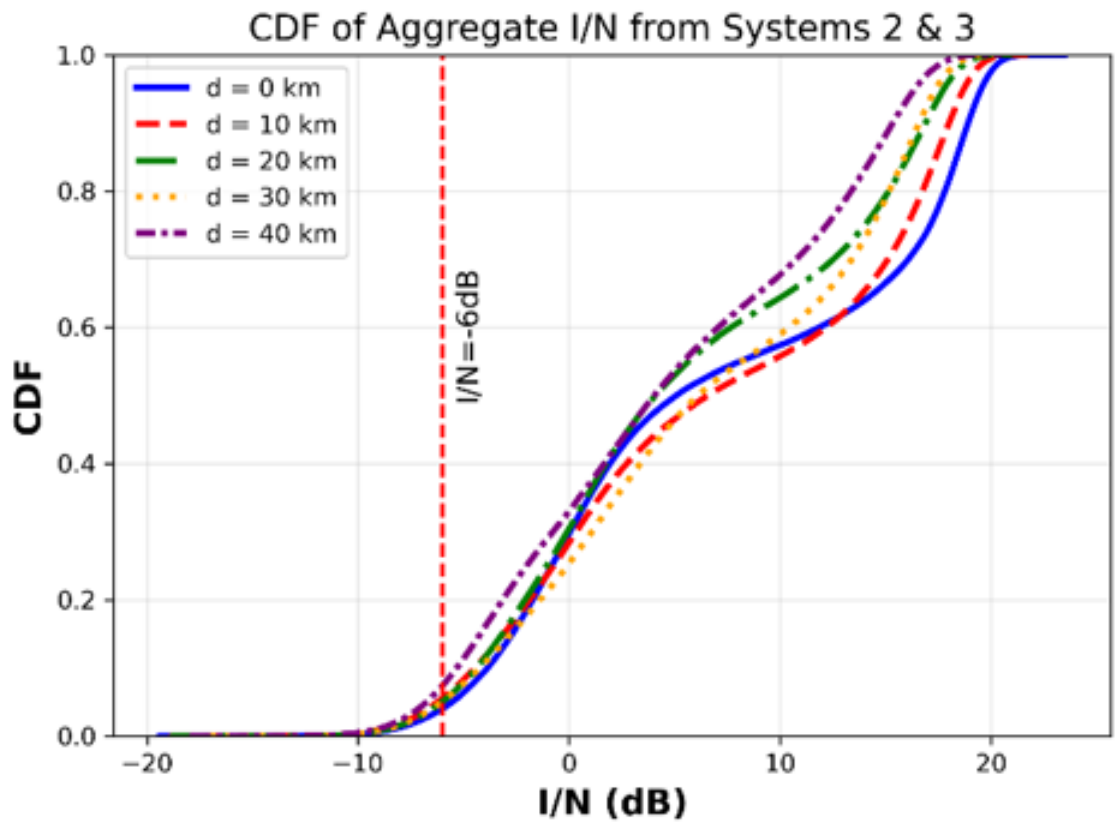

(b)

**FIGURE 8. The CDF of the aggregate I/N ratio of both D2D systems 2 & 3 at different distances from the border region of country A for (a) highest elevation satellite selection strategy and (b) random satellite selection strategy.**

For 80% of the time, at a strip distance of 0km (i.e., near the border of country A), the I/N ratio exceeds by 20dB, while at a strip distance of 40km, it exceeds by 13dB than the interference protection criteria for the highest elevation satellite selection strategy. However, in the case of a random satellite selection strategy, the I/N ratio exceeds by 24dB and 19dB at a strip distance of 0km and 40km, respectively. The I/N ratio and additional isolation at different time percentiles are listed in Table 5. The additional isolation is the amount, in dB, by which the interference power at the victim UE must be reduced in order for the aggregate I/N ratio to meet the -6dB protection criterion. To satisfy the long-term protection criteria of I/N=-6dB for 80% and 100% of CDF, additional isolation of up to 24dB and 27dB is needed, respectively. The random satellite selection strategy produces more interference compared to the highest elevating satellite selection strategy. This is because random satellite selection results in more cells in countries B and C being served by satellites at lower elevation angles. These low-elevation satellites direct their main beams at shallow angles toward the horizon, placing victims in Country A (Saudi Arabia) much closer to the main beam boresight, which significantly reduces the off-axis angle. As a result, these victims fall within the main lobe of the satellite antenna radiation pattern, where antenna gain is substantially higher compared to the deep sidelobe region encountered when using the highest elevation satellites. This combination of smaller off-axis angles and higher antenna gain leads to increased interference power reaching the victim, ultimately producing higher aggregate I/N levels.

**TABLE 5. The I/N ratio and required additional isolation (interference power reduction) at different time percentiles.**

| Satellite Selection Strategy | Distance | I/N (dB) | | Additional Isolation (I/N + 6dB) | |
|---|---|---|---|---|---|
| | | 80% CDF | 100% CDF | 80% CDF | 100% CDF |
| Highest Elevation Selection Strategy | 0 km | 14 | 17 | 20 | 23 |
| | 40 km | 7 | 8 | 13 | 14 |
| Random Selection Strategy | 0 km | 18 | 21 | 24 | 27 |
| | 40 km | 13 | 18 | 19 | 24 |

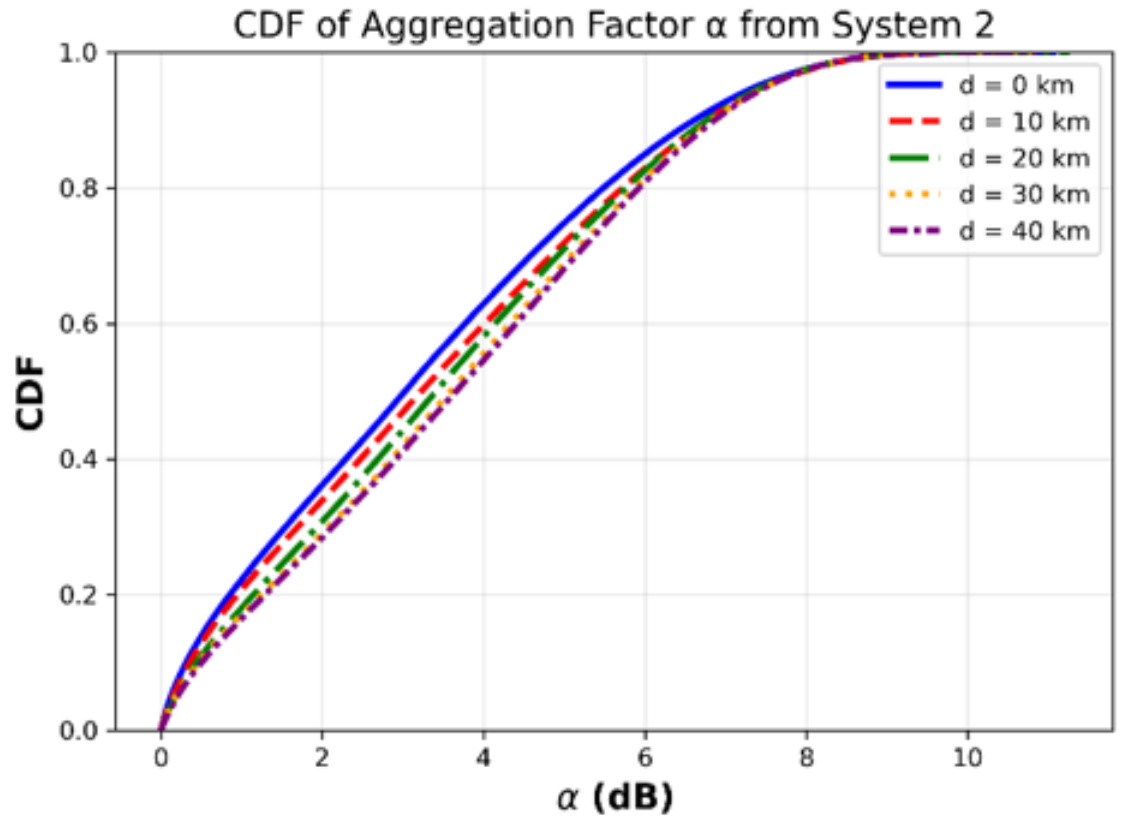

(a)

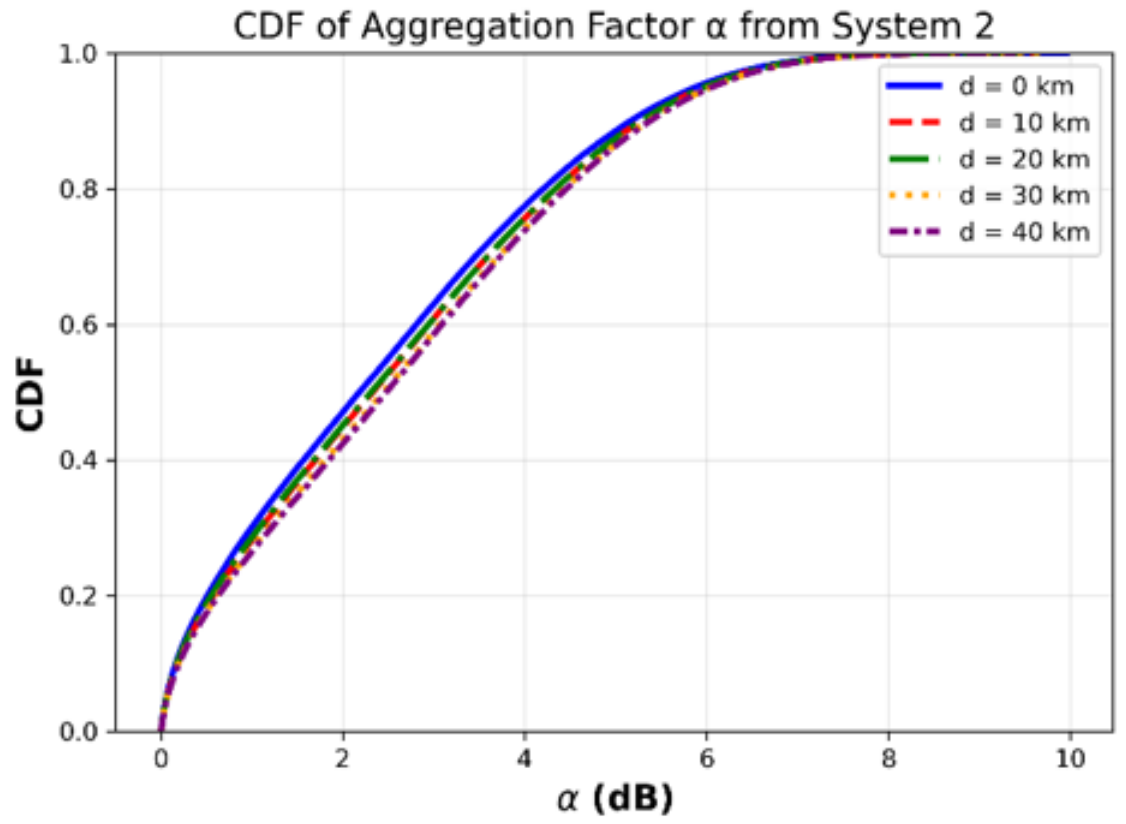

(b)

**FIGURE 9. The CDF of the aggregation factor α of the D2D System 2 at different distances from the border region of country A for (a) highest elevation satellite selection strategy and (b) random satellite selection strategy.**

### B. INTRA-SYSTEM AND INTER-SYSTEM AGGREGATION FACTORS

The CDF of the intra-system aggregation factor α is plotted in Figs. 9 and 10 for Systems 2 and 3, respectively. The aggregation factor α increases marginally with strip distance, representing a negligible change in the number of contributed satellites. The results show that the aggregation factor α reaches up to approximately 8.5dB for both systems, indicating that multiple satellites contribute comparably to the aggregate interference under worst-case conditions. The CDF of the inter-system aggregation factor β is shown in Fig. 11 for different strip distances from the border of country A under both satellite selection strategies. The impact of victim location on the aggregation factor β is small. The factor (β) reaches a value up to 3dB and is small for most of the distribution (e.g., β ~ 0.5dB at 80% of CDF), confirming that one D2D system dominates the total interference, i.e., System 3 in this study. The β distributions for the five strips are essentially indistinguishable under both selection strategies, indicating that the relative apportionment of interference between the two systems does not change significantly with distance from the border.

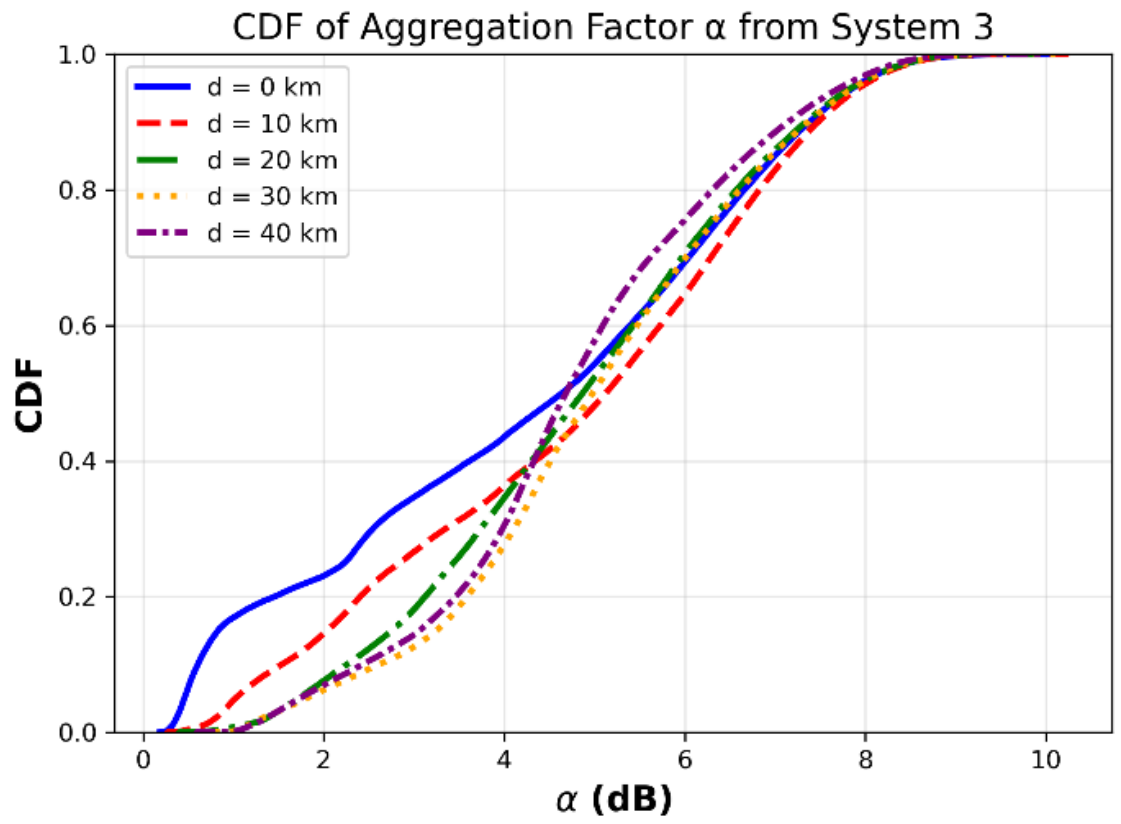

(a)

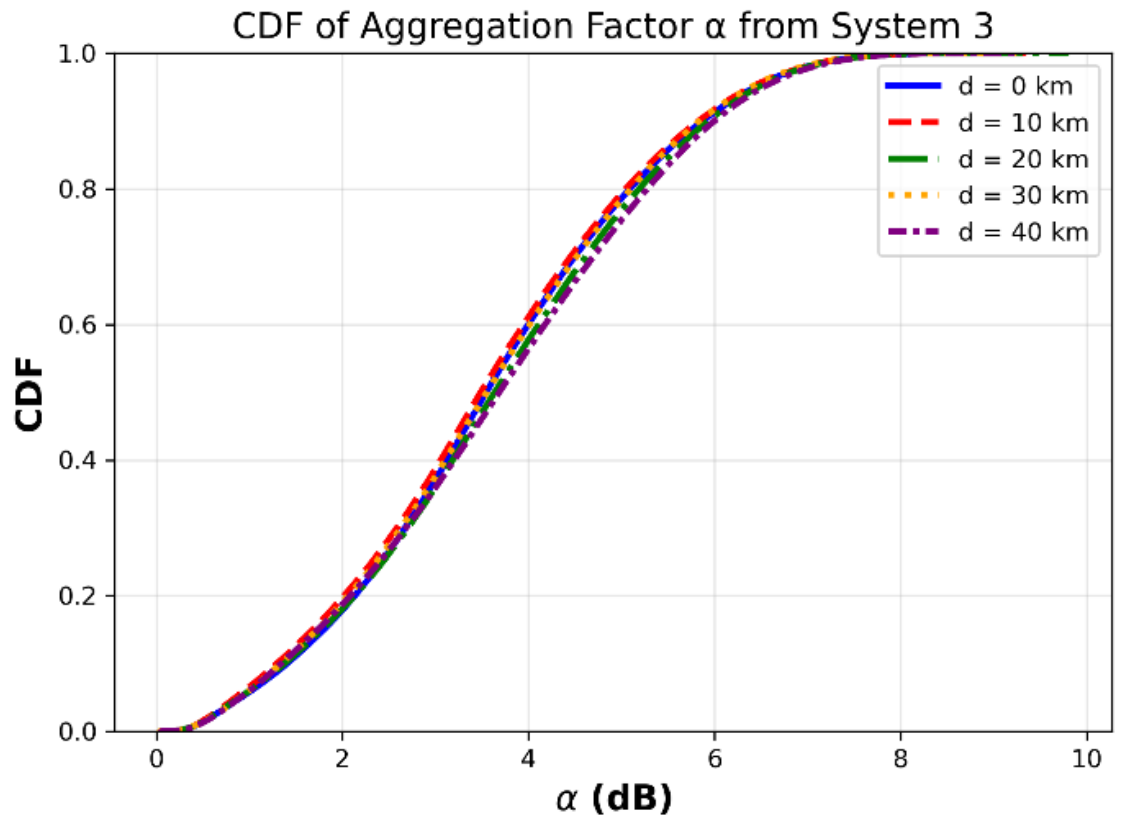

(b)

**FIGURE 10. The CDF of the aggregation factor α of the D2D System 3 at different distances from the border region of country A for (a) highest elevation satellite selection strategy and (b) random satellite selection strategy.**

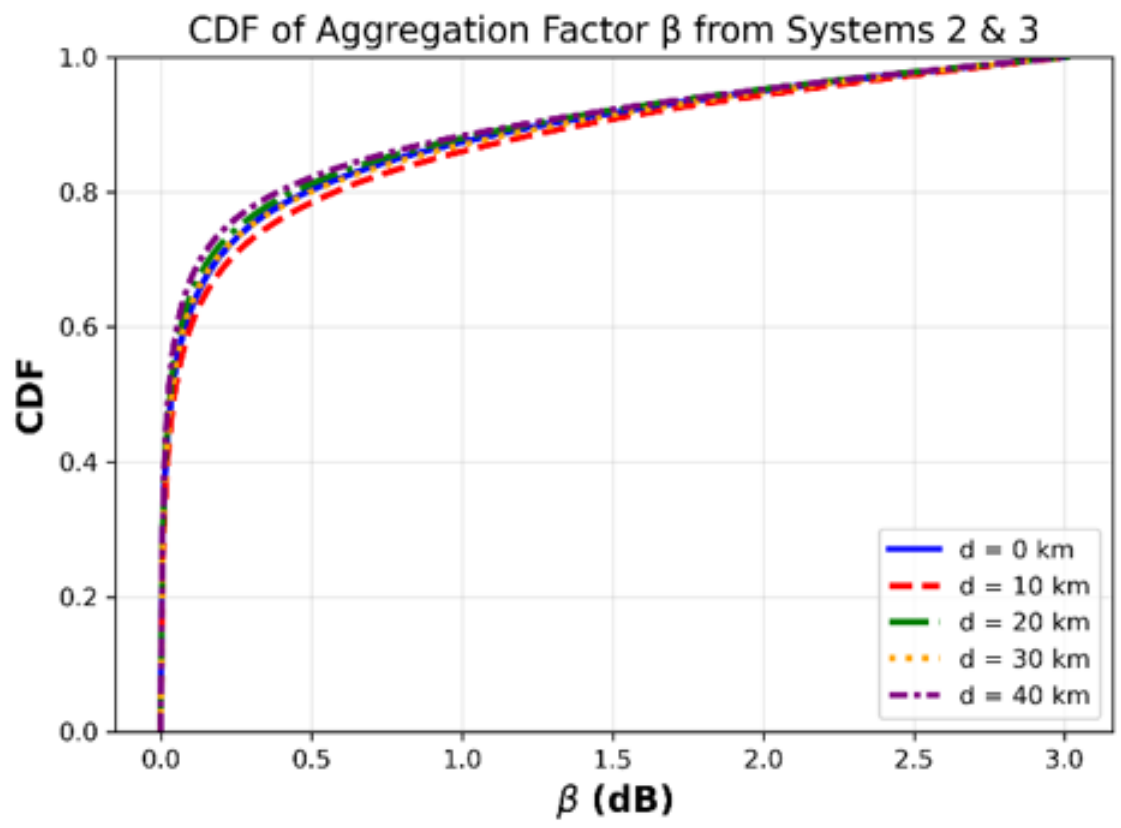

(a)

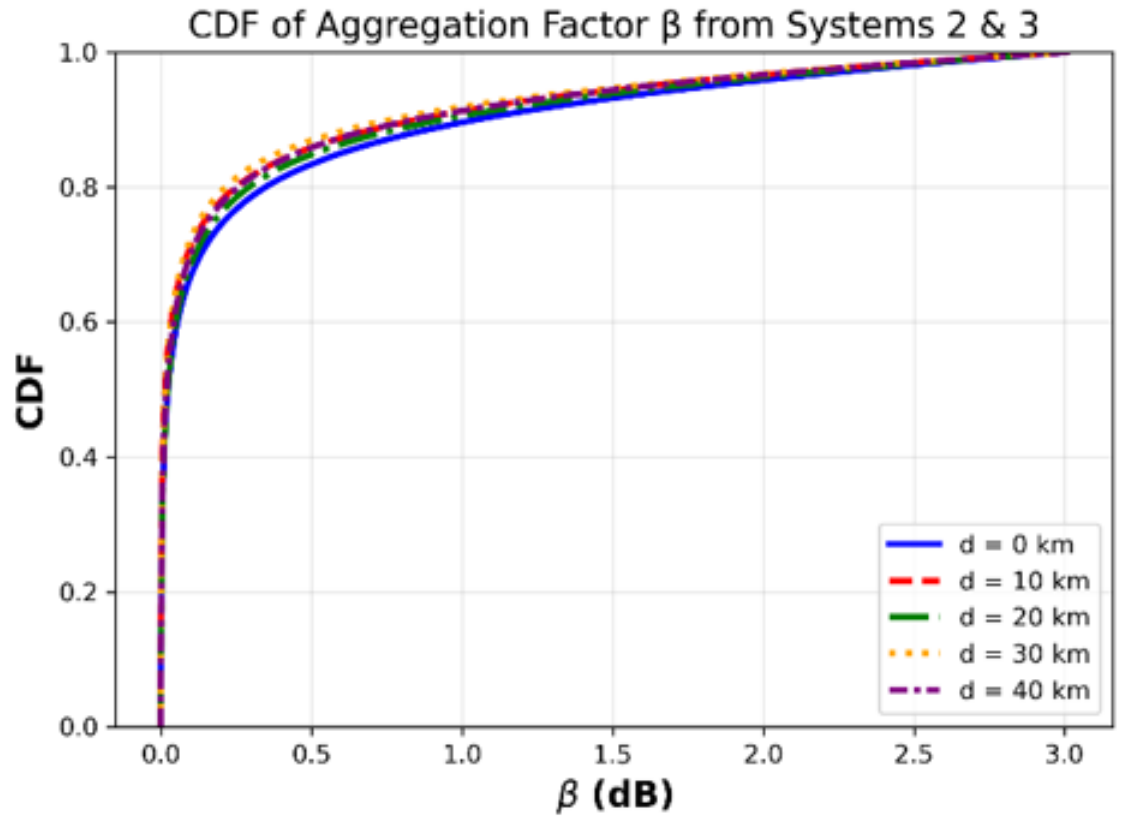

(b)

**FIGURE 11. The CDF of the aggregation factor β of both D2D Systems 2 & 3 at different distances from the border region of country A for (a) highest elevation satellite selection strategy and (b) random satellite selection strategy.**

## *C. CELL-LEVEL INTERFERENCE ANALYSIS*

To better understand the impact of the D2D Systems on interference affecting victim UEs within Saudi Arabia, as well as potential mitigation techniques, we analyzed interference from a sample of System 3 satellites. These satellites serve a cell nearest to the border of Country A (Cell 3) and a cell farthest from it (Cell 6), as illustrated in Fig. 12. Under the highest-elevation satellite selection technique, Figs. 13 and 14 show that the I/N ratio for satellites covering Cell 3 exceeds the protection criterion, whereas interference from satellites covering Cell 6 remains within acceptable limits. This indicates that when satellites direct their beams at the highest elevation angles toward the cells, those serving the areas closest to the border of Country A become the dominant source of interference. In these cases, victim UEs are more likely to fall within the satellite's main beam and experience lower path loss due to shorter propagation distances. In contrast, under the random satellite selection strategy, the I/N ratio exceeds the protection criterion for satellites covering both the nearest and farthest cells. This result shows that even satellites serving areas farther from the border can still cause significant interference to victims in Country A. Because these randomly selected satellites may transmit at lower elevation angles, victim UEs can fall within the main beam, leading to high interference levels. We analyzed interference caused by satellites serving other nearest and farthest cells to the border, and we observed similar behavior; the corresponding results are omitted here to avoid redundancy.

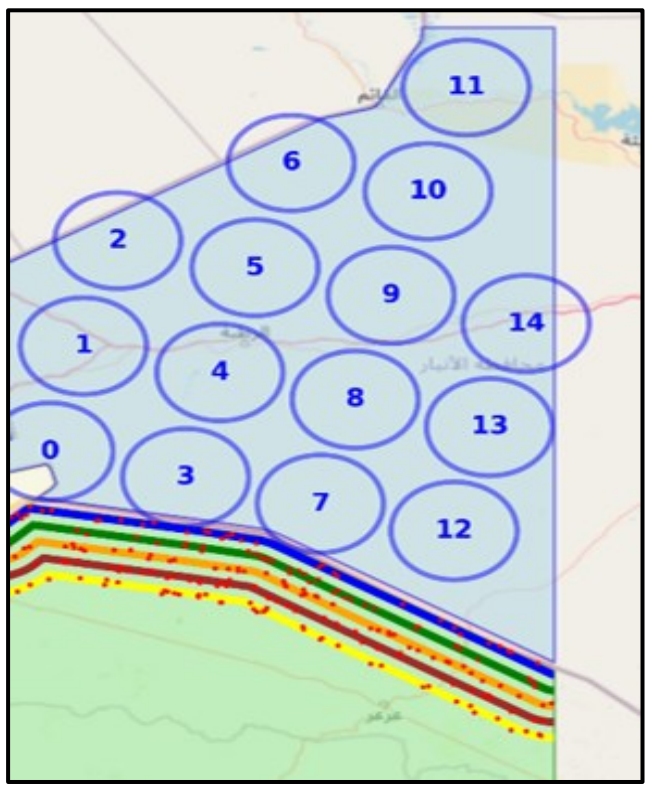

**FIGURE 12. The cell arrangement of the D2D System 3 in country C (Iraq): Cell 3 is located near the border, while Cell 6 is located far from the border of country A (Saudi Arabia).**

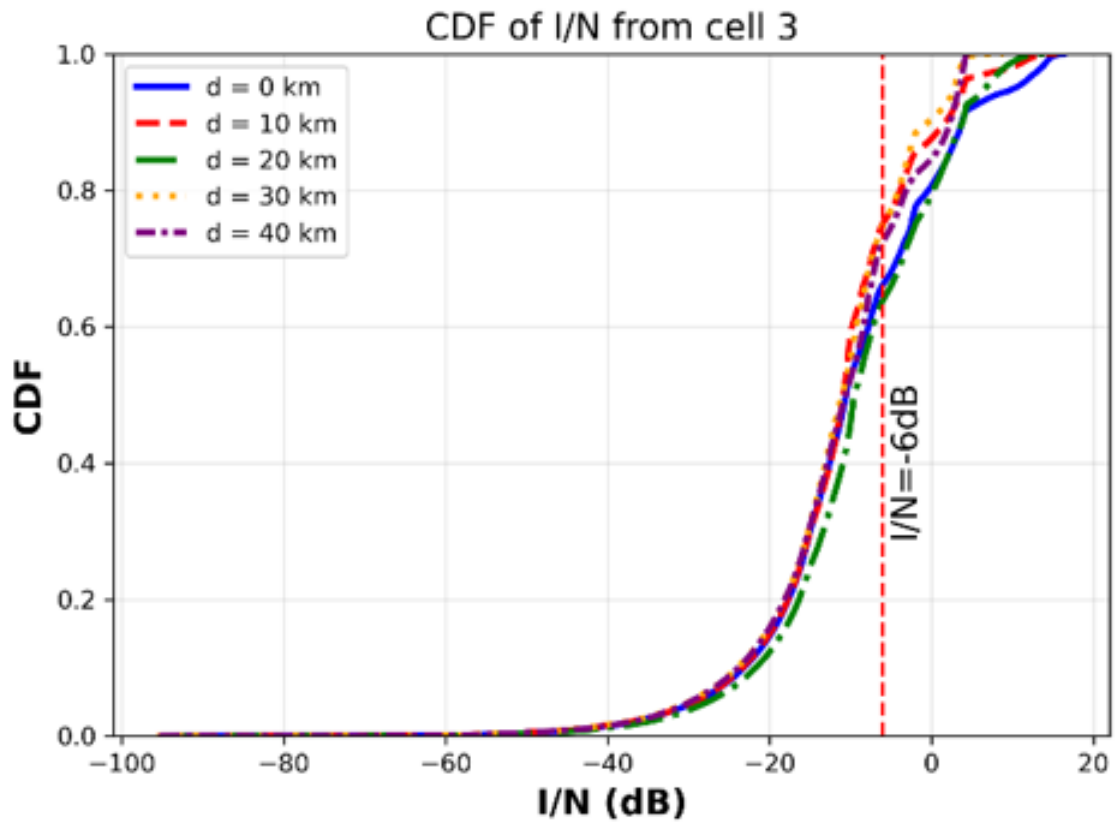

(a)

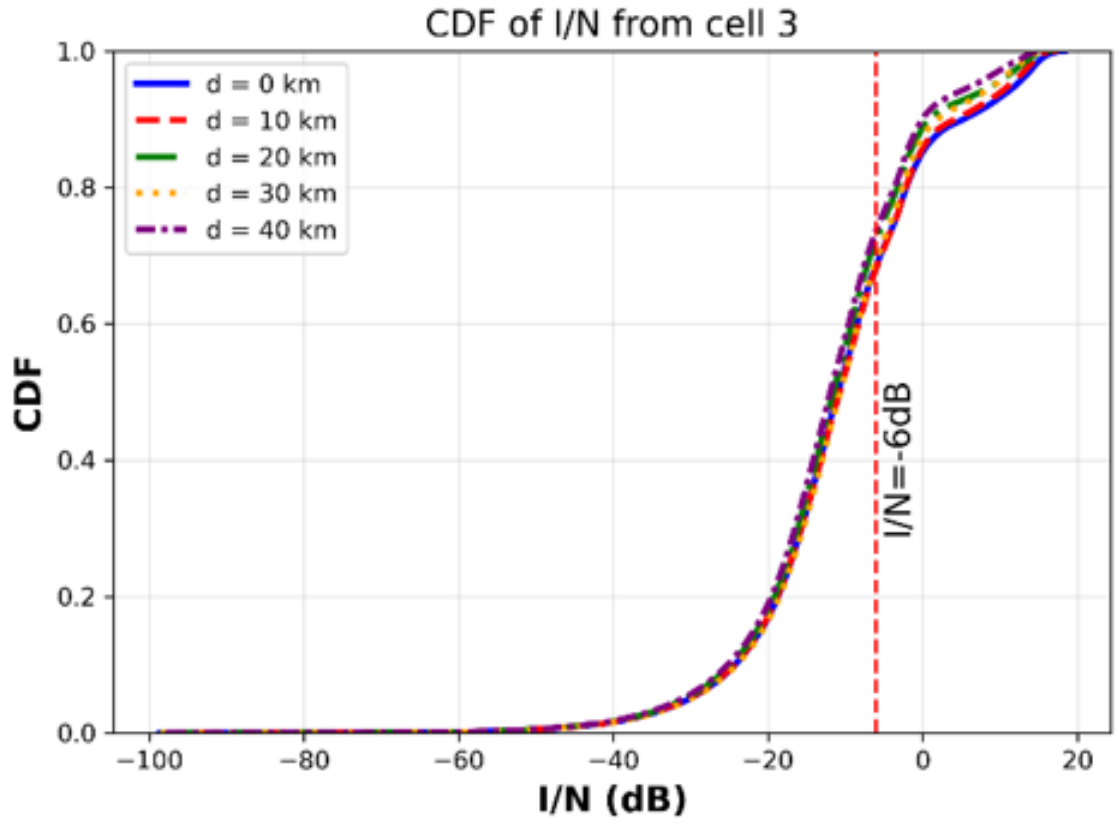

(b)

**FIGURE 13. The CDF of the I/N ratio for satellites covering Cell 3 of the D2D System 3, located near the country A border for (a) highest elevation satellite selection strategy and (b) random satellite selection strategy.**

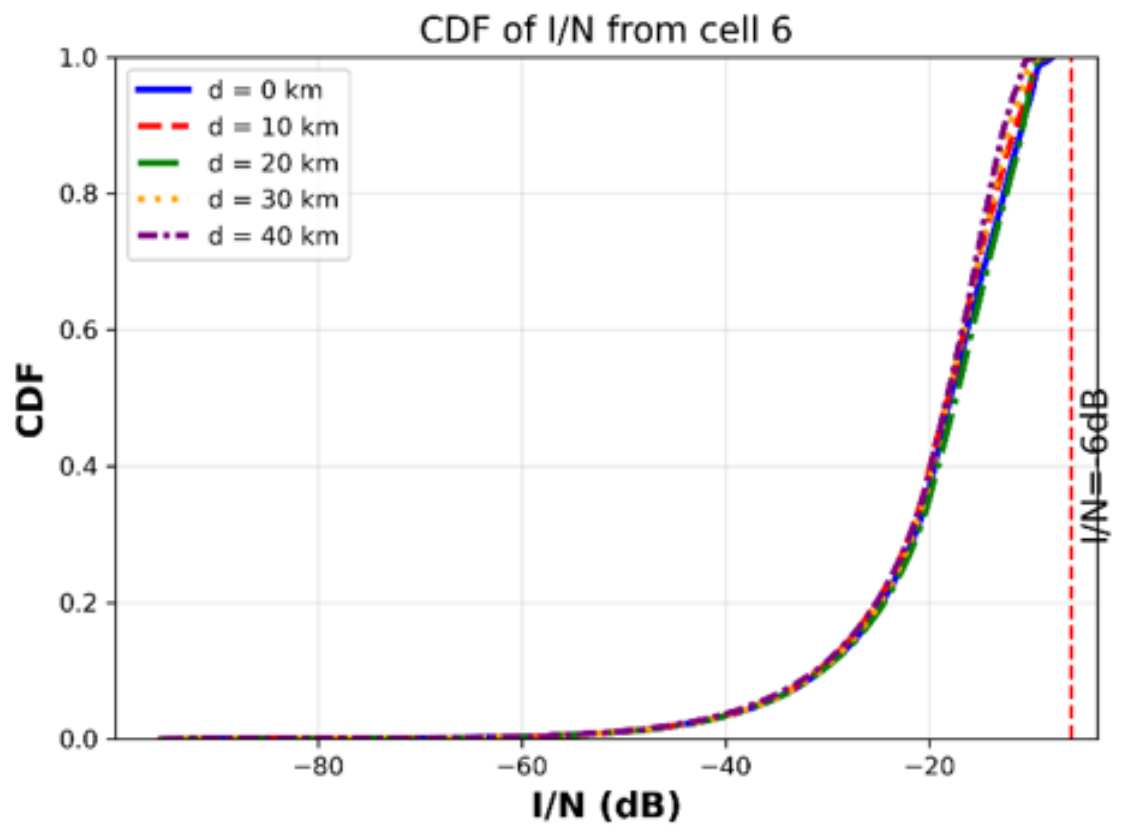

(a)

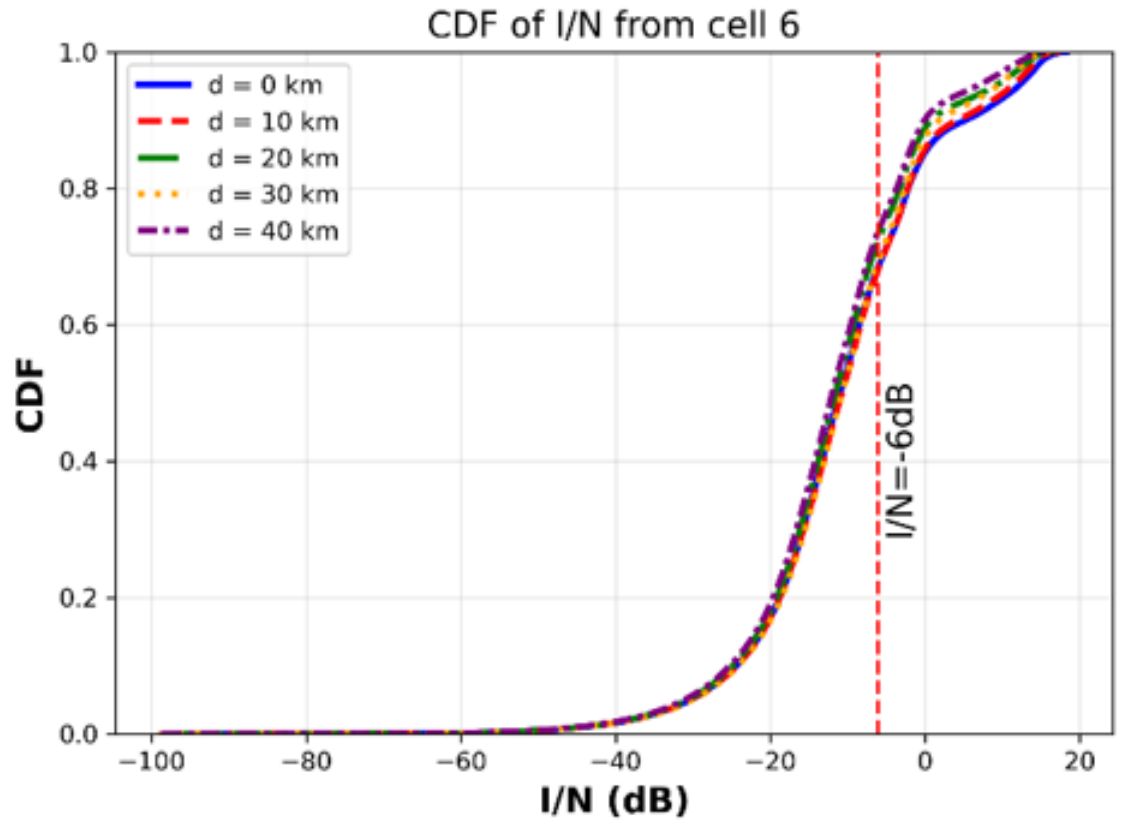

(b)

**FIGURE 14. The CDF of the I/N ratio for satellites covering Cell 6 of the D2D System 3, located far from the country A border for (a) highest elevation satellite selection strategy and (b) random satellite selection strategy.**

## *D. INTERFERENCE MITIGATION TECHNIQUES*

As a mitigation technique, we examine the effect of introducing an isolation distance equal to the cell diameter of each neighboring D2D system along the border of countries B and C, i.e., excluding ground cells closest to the border region from D2D service. To avoid redundant figures, Fig. 15 shows the CDF of the aggregate I/N ratio for the combined D2D Systems 2 and 3 after applying this isolation distance; the CDFs for the individual systems follow the same trend and are therefore omitted. Under the highest-elevation selection strategy, the isolation distance reduces the exceedance of the protection criterion from more than 80% (Fig. 8(a)) to roughly 61-73%. Under the random selection strategy, the exceedance falls from about 90-96% (Fig. 8(b)) to approximately 80-82%. The isolation zone therefore provides a moderate and broadly comparable improvement under both selection strategies, rather than favoring one strategy over the other. In both cases, the dependence of the exceedance on the strip distance is weak and not monotonic, so that moving the victim farther from the border does not systematically improve compliance once the border-adjacent cells have been removed from service. This is because excluding those cells removes only the shortest-range, smallest off-axis-angle contributions; the remaining interference is dominated by satellites serving cells deeper inside countries B and C, whose slant paths to the victim vary little across the 40km span of the victim strips. Most importantly, the aggregate I/N ratio still exceeds the -6dB criterion for most of the time under both strategies, so an isolation distance of one cell diameter is not sufficient to achieve coexistence.

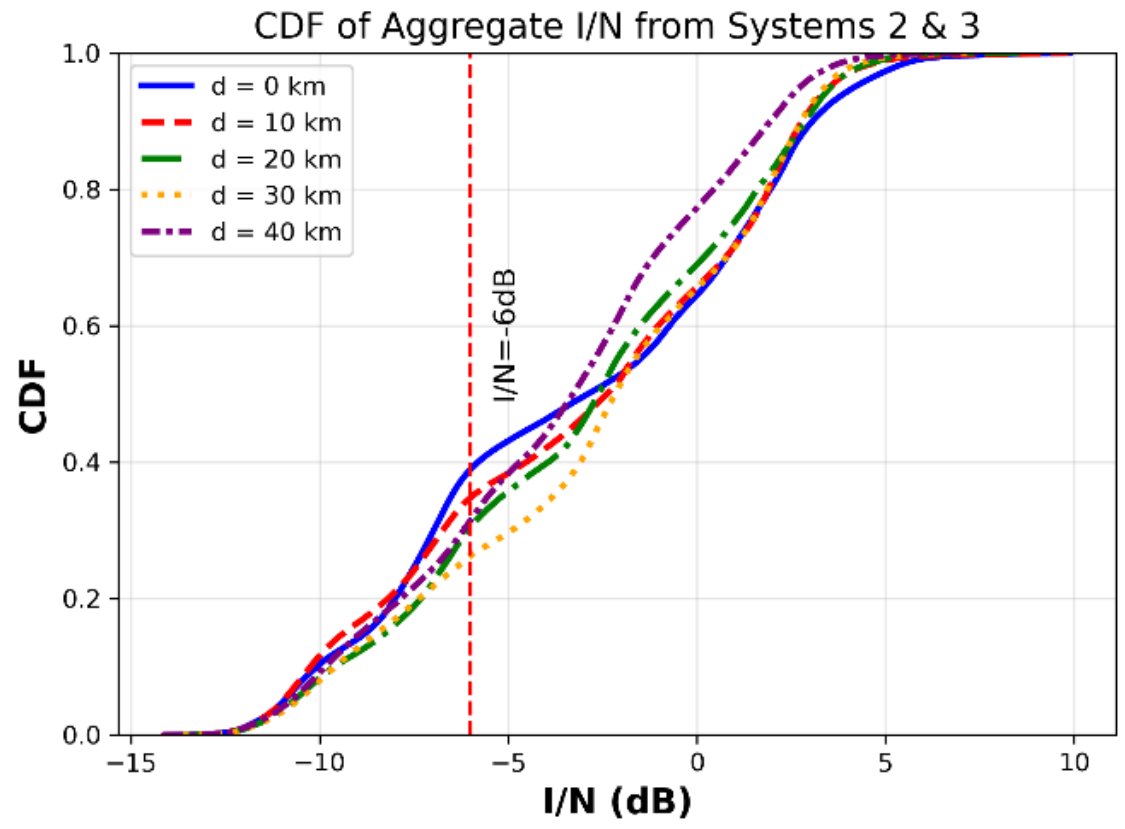

(a)

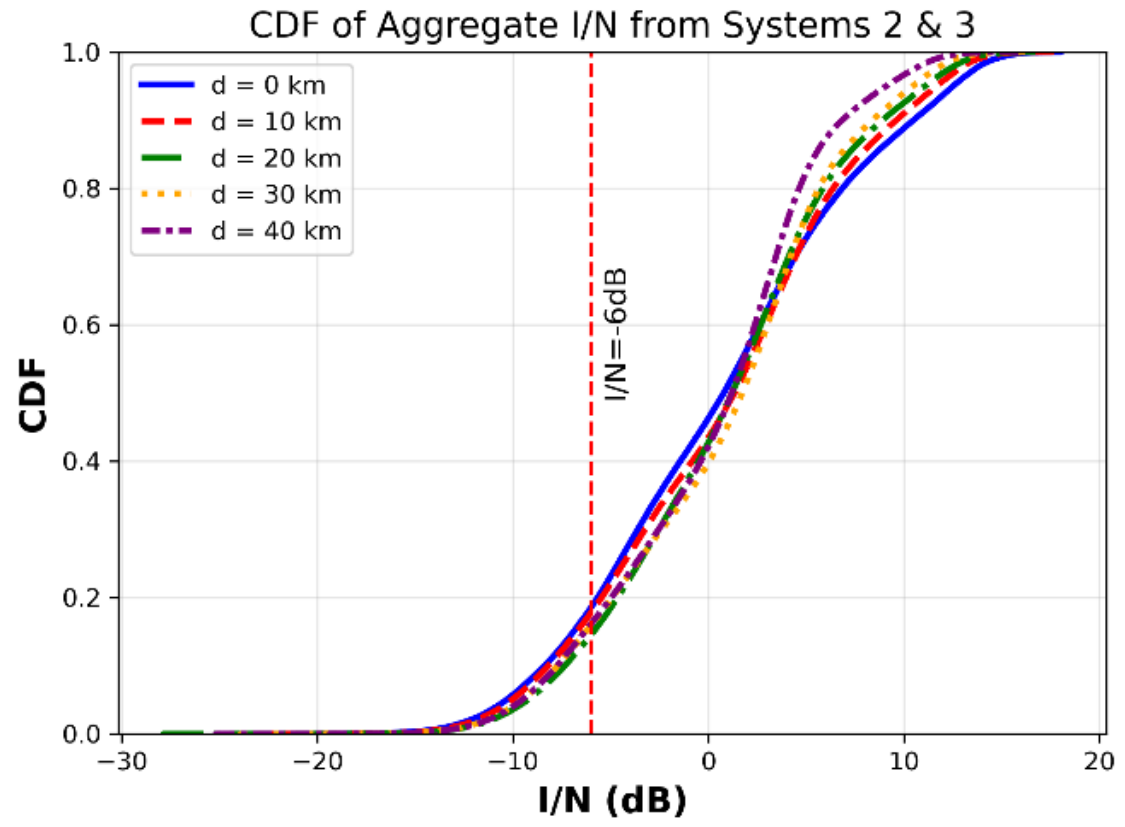

(b)

**FIGURE 15. The CDF of the aggregate I/N ratio of both D2D Systems 2 & 3 at different distances from the border region of country A after applying an isolation distance equal to the cell diameters of each D2D system in the neighboring countries for (a) the highest elevation satellite selection strategy and (b) the random satellite selection strategy.**

As another mitigation approach, we replace the antenna radiation pattern of System 3 with the phased-array pattern used by System 2 (Rec. ITU-R M.2101), while keeping all other orbital and link-budget parameters unchanged. This tests the hypothesis that the narrower beam and deeper sidelobe nulls of the phased array pattern are responsible for the lower interference produced by System 2. Figure 16 shows the resulting CDF of the aggregate I/N ratio from both D2D systems. Adopting the phased array pattern for System 3 reduces the exceedance of the protection criterion to approximately 40%, compared with about 90% in the baseline case of Fig. 8. The stepped shape of the baseline System 3 curves is likewise removed, confirming that it is produced by the discrete sidelobe structure of the ITU-R S.1528 radiation pattern. This confirms that the antenna radiation pattern, rather than the satellite selection strategy or the isolation distance, is the dominant factor behind the excess interference contributed by System 3. This requires D2D systems operating in these bands to adopt phased-array antennas with deep sidelobe nulls, which is a substantially more effective and strategy-independent mitigation technique than a border isolation zone alone. Nevertheless, even with this improved antenna pattern, the aggregate I/N ratio still exceeds the -6dB protection criterion for roughly one third of the time, indicating that antenna pattern improvement alone is not sufficient to guarantee full compliance and should be combined with additional measures, such as power control, beam exclusion zones, or elevation-angle masking, to fully satisfy the IMT protection criterion.

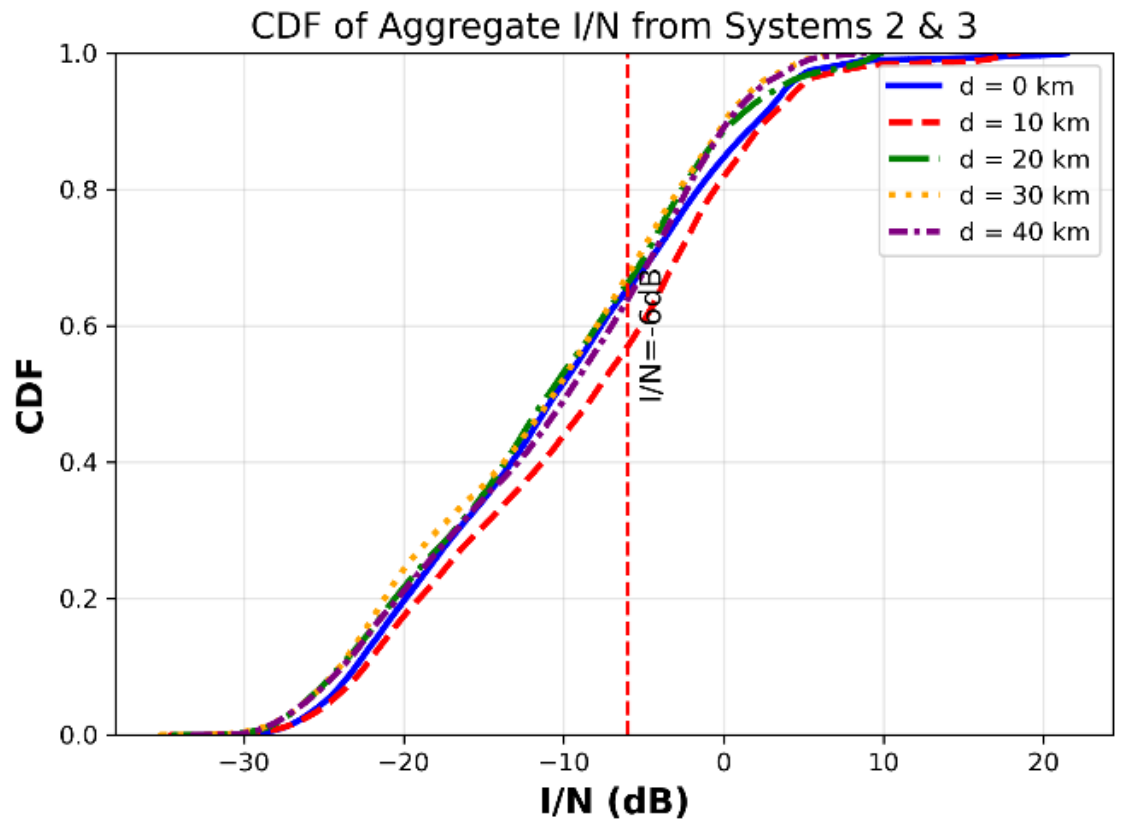

(a)

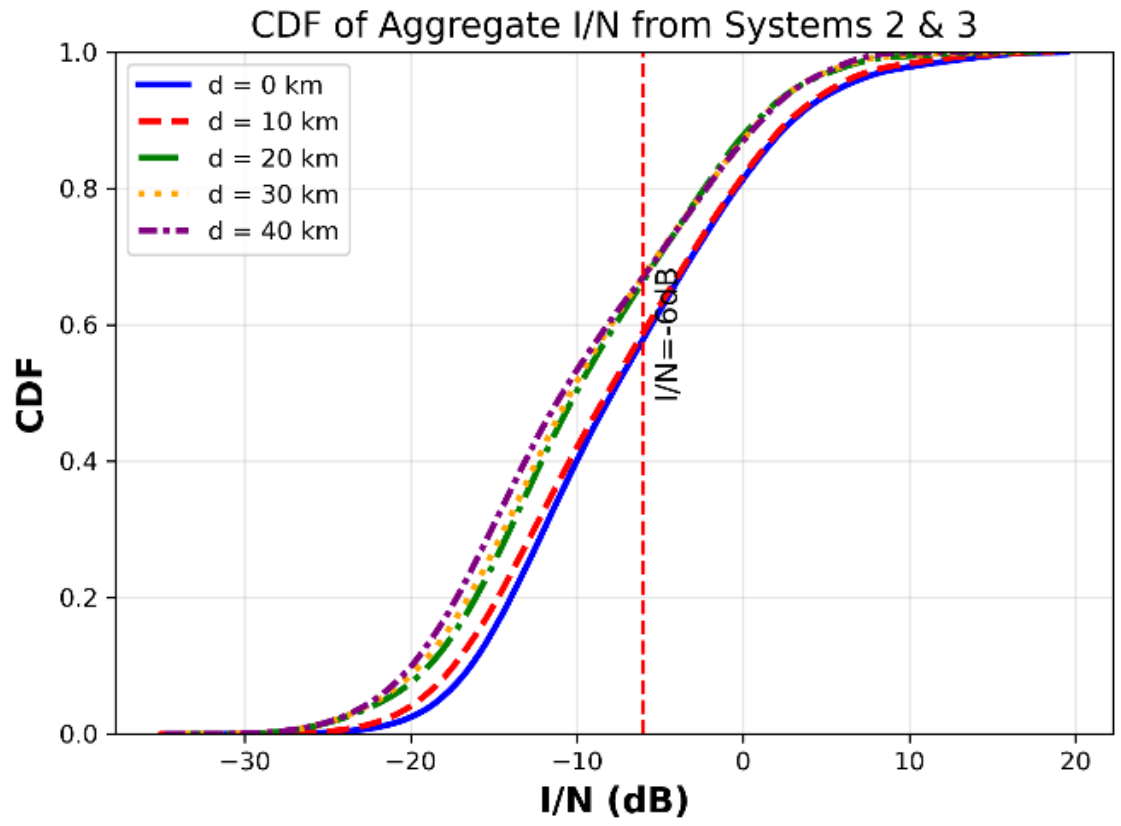

(b)

**FIGURE 16. The CDF of the aggregate I/N ratio of both D2D systems 2 & 3 at different distances from the border region of country A after using the ITU-R M.2101 phased-array antenna radiation pattern in both D2D systems for (a) highest elevation satellite selection strategy and (b) random satellite selection strategy.**

Overall, the results in this section show that the cross-border aggregate interference from D2D Systems 2 and 3 into terrestrial IMT UEs in Saudi Arabia significantly exceeds the -6dB protection criterion under baseline conditions, with System 3 acting as the dominant interference source owing to its wider-beam and shallower-null antenna pattern. Among the two mitigation techniques examined, a fixed geographic isolation distance of one cell diameter provides only a moderate reduction in the exceedance, of broadly similar magnitude under both satellite selection strategies, and leaves the aggregate I/N ratio above the criterion for most of the time. In contrast, adopting a narrower-beam, deep-null antenna pattern like Rec. ITU-R M.2101 for System 3 is considerably more effective, roughly halving the exceedance of the protection criterion and making the results largely insensitive to the satellite selection strategy. These findings suggest that, from a regulatory viewpoint, antenna pattern requirements may offer a more robust and strategy-independent basis for mitigating cross-border D2D interference than geographic isolation zones alone. In addition, combining these techniques with additional measures such as power control or elevation-angle masking would likely be needed to fully satisfy the IMT protection criterion along the border region of Saudi Arabia, Jordan, and Iraq.

## V. CONCLUSION AND FUTURE WORK

This study assessed the cross-border aggregate downlink interference from two D2D systems, operating in Jordan and Iraq, into terrestrial IMT UEs in Saudi Arabia in the 698-960 MHz band. In this study, the D2D System 2 is a LEO satellite constellation with a 500 km altitude and a 35° minimum elevation angle, assumed to provide service coverage in Jordan, while System 3 operates at 525 km altitude with a 20° minimum elevation angle, assumed to provide service coverage in Iraq. System 3 exceeds the protection criterion up to 88% of the time and System 2 up to 18%, requiring up to 27 dB of additional isolation, so coexistence is not possible without interference mitigation. Several satellites of a single dominant D2D system drive aggregate interference. This is evidenced by the intra-system factor reaching about 8.5 dB, while the inter-system factor stays below 3 dB.

The satellite selection strategy strongly drives cross-border interference. The highest-elevation strategy confines harmful interference mainly to satellites serving border-proximate cells, whereas random selection places victims near the boresight of low-elevation beams from distant cells and yields higher interference. The results illustrate that an isolation zone of one cell diameter used as a mitigation technique gives only a moderate reduction under both strategies. However, a narrow-beam, deep-null phased-array pattern for System 3 lowers the exceedance to about 40% and makes it largely independent of selection strategy. Antenna pattern requirements are therefore more robust regulatory control than isolation alone, though neither achieves full compliance. Future work will extend the framework to IMT base stations as victims, assess realistic operational assumptions such as partial load and power control, consider other IMT bands, and refine the propagation model with clutter and terrain effects.

## ACKNOWLEDGMENT

The authors would like to thank the leadership of radio spectrum in the Communications, Space & Technology Commission (CST) of Saudi Arabia for their support of this work.

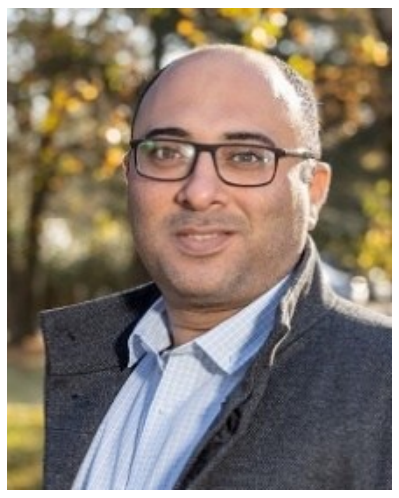

**Muneer M. Al-ZuBi** received the Ph.D. degree in engineering from the University of Technology Sydney (UTS), Sydney, Australia, in 2020. He worked as a Research Associate with the Center of Excellence for Innovative Projects, Jordan University of Science and Technology (JUST). He was a postdoctoral researcher with the Department of Engineering, University of Luxembourg, Luxembourg, and also, he was a remote visiting scholar with the School of Electrical and Data Engineering, UTS. He is currently a Postdoctoral Fellow at the Communication Theory Lab (CTL), King Abdullah University of Science and Technology (KAUST), Saudi Arabia. His research interests lie in the areas of wireless communication, radio wave propagation, and molecular communication.

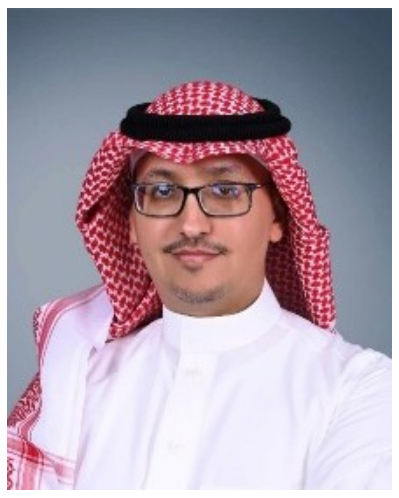

**Muhammad Alrabeiah** is an assistant professor and a subject-matter expert specializing in wireless communications. He is currently a full-time advisor to the spectrum sector of the Communication, Space, and Technology Commission (CST) of Saudi Arabia. He serves as a member of the national Spectrum Advisory Group (SAG) of Saudi Arabia. In addition, he holds an assistant professor position in the faculty of engineering at King Saud University (KSU), Riyadh, Saudi Arabia. Dr. Alrabeiah received his PhD in electrical engineering in 2021 from the School of Electrical, Computer, and Energy Engineering (ECEE) of Arizona State University (ASU), Tempe, AZ, USA, and he received his Master of Applied Sciences (MAS) in 2016 from the Department of Electrical and Computer Engineering (ECE) at McMaster University in Hamilton, ON, Canada. Dr. Alrabeiah's professional expertise and research interests span the fields of machine learning, wireless technologies, and telecommunications and radio spectrum policy.

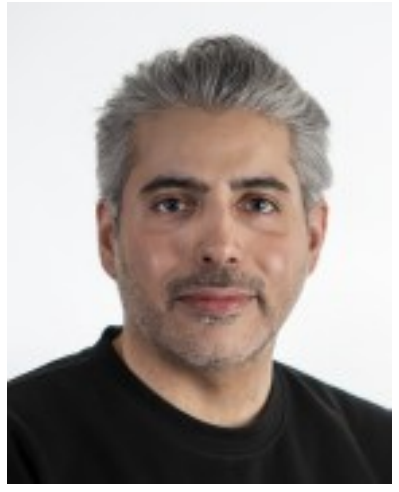

**Khalid Aldubaikhy** (Member, IEEE) is an Associate Professor with the Department of Electrical Engineering at Qassim University, Saudi Arabia, where he also serves as Advisor to the Vice President for Postgraduate Studies and Scientific Research. He received the B.E. degree in electrical engineering from Qassim University in 2008, the M.A.Sc. degree in electrical and computer engineering from Dalhousie University, Halifax, NS, Canada, in 2012, and the Ph.D. degree in electrical and computer engineering from the University of Waterloo, Waterloo, ON, Canada, in 2019. His research interests include millimeter-wave and terahertz wireless networks, non-terrestrial networks (NTNs), spectrum management, low Earth orbit (LEO) satellite systems, and the application of artificial intelligence in next-generation communication systems.

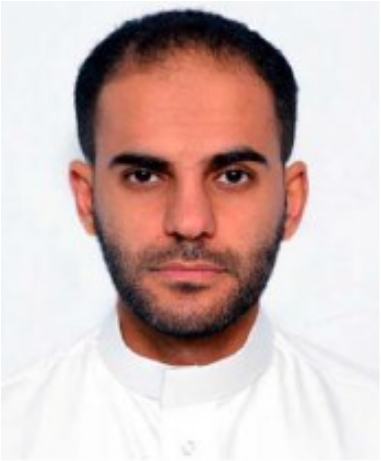

**Abdullah Alqasir** (Member, IEEE) received the B.S. degree from Qassim University, Buraydah, Saudi Arabia, in 2008, the M.S. degree from The Ohio State University, Columbus, OH, USA, in 2014, and the Ph.D. degree in electrical and computer engineering from Iowa State University, Ames, IA, USA, in 2019. He is currently an Associate Professor with the Department of Electrical Engineering, Qassim University, Buraydah, Qassim, Saudi Arabia. His research interests include satellite communication, non-terrestrial networks (NTN), next-generation networks, 6G, and ISAC.

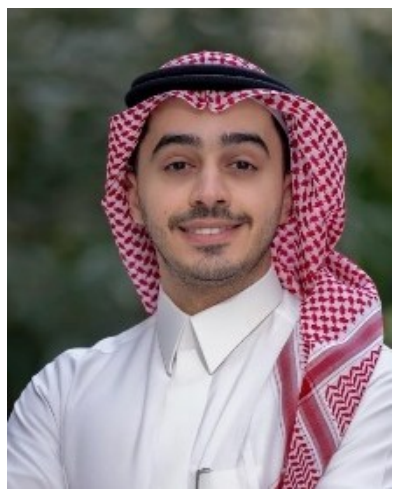

**Majed Alsifaee** received the B.Sc. degree in electrical engineering from Imam Mohammad Ibn Saud Islamic University, Riyadh, Saudi Arabia, in 2018, and the M.Sc. degree in finance from King Saud University (KSU), Riyadh, Saudi Arabia, in 2023. He is currently a Spectrum Management Expert with the Communications, Space and Technology Commission (CST), Saudi Arabia, working on radio spectrum management, licensing, and regulatory frameworks. He is a member of the Saudi national team for ITU World Radiocommunication Conference (WRC) preparations and represents Saudi Arabia in ITU radiocommunication activities, including ITU Study Group WP4C. He also leads spectrum auction initiatives and contributes to national spectrum strategy development. His research interests include spectrum management, spectrum sharing, mobile and satellite communications, international spectrum regulation, and emerging wireless technologies.

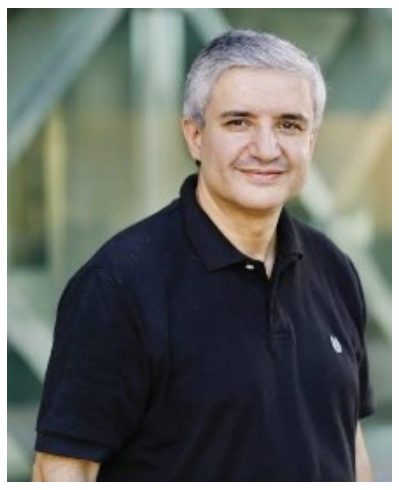

**Mohamed-Slim Alouini** [S'94, M'98, SM'03, F'09] received the Ph.D. degree in electrical engineering from the California Institute of Technology (Caltech), Pasadena, CA, USA, in 1998. He served as a faculty member at the University of Minnesota, Minneapolis, MN, USA, then at Texas A&M University at Qatar, Education City, Doha, Qatar, before joining King Abdullah University of Science and Technology (KAUST), Thuwal, Makkah Province, Saudi Arabia, as a professor of electrical engineering in 2009. His current research interests include modeling, design, and performance analysis of wireless communication systems.